
\magnification 1200
\input amstex
\documentstyle{amsppt}

\font\bigtenrm=cmr10 scaled\magstep2
\def\non{{{\noindent}}}
\def\a{\alpha}
\def\ung{{{\frak{g}}}}

\def\uqg{{{Y(\ung)}}}

\def\ot{{{\otimes}}}
\def\op{{{\oplus}}}

\def\unh{{{\frak{h}}}}

\def\ungj{\ung_J}

\def\uqgj{Y(\ung_J)}

\def\non{{{\noindent}}}
\def\ung{{{\frak{g}}}}

\def\uqg{{{Y(\ung)}}}

\def\unk{\frak k}

\def\ot{{{\otimes}}}
\def\op{{{\oplus}}}

\def\unh{{{\frak{h}}}}

\def\ungj{\ung_J}

\def\uqgj{Y(\ung_J)}

\def\ung{\frak g}
\def\unh{\frak h}
\def\op{\oplus}
\def\y{{Y(\ung)}}
\def\ot{\otimes}
\def\l{\lambda}
\def\g{\gamma}
\def\k{\kappa}
\def\non{{{\noindent}}}
\def\ung{{{\frak{g}}}}

\def\uqg{{{Y(\ung)}}}

\def\unk{\frak k}
\def\ot{{{\otimes}}}
\def\op{{{\oplus}}}

\def\unh{{{\frak{h}}}}

\def\ungj{\ung_J}

\def\uqgj{Y(\ung_J)}
\def\v{\varphi}

\def\non{{{\noindent}}}
\def\ung{{{\frak{g}}}}

\def\uqg{{{Y(\ung)}}}

\def\ot{{{\otimes}}}
\def\op{{{\oplus}}}

\def\unh{{{\frak{h}}}}

\def\ungj{\ung_J}

\def\uqgj{Y(\ung_J)}

\NoBlackBoxes
\nologo
\document
\centerline{\bigtenrm{\bf Yangians, Integrable Quantum Systems and Dorey's
Rule}}
\vskip18pt\centerline{Vyjayanthi Chari and Andrew Pressley}
\vskip36pt
{\eightpoint{
\hskip3.5cm 1. Introduction

\hskip3.5cm 2. Yangians

\hskip3.5cm 3. Finite dimensional representations

\hskip3.5cm 4. Dorey's rule

\hskip3.5cm 5. Some preliminary lemmas

\hskip3.5cm 6. The $A_n$ case

\hskip3.5cm 7. The $D_n$ case

\hskip3.5cm 8. The $B_n$ and $C_n$ cases

\hskip3.5cm 9. The quantum affine case

\hskip3.5cm 10. Appendix: Dorey's rule and affine Toda theories

\hskip3.5cm 11. References
}}
\vskip36pt\centerline{\bf 1. Introduction}
\vskip12pt\noindent Quantum groups arose from the quantum inverse scattering
method, developed by the Leningrad school [13] to solve integrable quantum
systems. They provide, in particular, a way to understand the solutions of the
quantum Yang--Baxter equation (R-matrices) associated to such systems, and a
general framework for producing new solutions. Of special importance are the
solutions which depend on a complex (\lq spectral') parameter; those which are
rational, or trigonometric, functions of this parameter arise from the quantum
groups called Yangians, or quantum affine algebras, respectively (see [11],
[12] and Chapter 12 in [8] for background information).

More recently, quantum groups have arisen in another guise in connection with
$1+1$ dimensional integrable quantum field theories, namely as the algebras
satisfied by certain non-local conserved currents. For example, Yangians appear
as \lq quantum symmetry algebras\rq\ in $G$-invariant Wess--Zumino--Witten
models [1], while quantum affine algebras appear in affine Toda field theories
(ATFTs) [2]. In [10], Dorey gave a remarkable Lie-theoretic description of the
classical three-point couplings (or \lq fusings\rq) in certain integrable field
theories, including ATFTs. It is the purpose of this paper to interpret Dorey's
rule in terms of the representation theory of Yangians and quantum affine
algebras.

\topspace{2cm}

To describe our results in more detail, recall that an ATFT is a theory of
scalar fields with exponential interactions determined by the roots of a
(possibly twisted) affine Lie algebra. If $\ung$ is a finite-dimensional
complex simple Lie algebra, and $\hat\ung$ is the associated (untwisted) affine
Lie algebra, the quantum affine algebra $U_\epsilon(\hat\ung)$ is a \lq quantum
symmetry algebra\rq\ of the ATFT based on the dual affine algebra
${\hat\ung}^*$, whose Dynkin diagram is obtained from that of $\hat\ung$ by
reversing the arrows (the deformation parameter $\epsilon$ is related to the
coupling constant of the theory, which should be purely imaginary for the
quantum affine symmetry to exist -- see Section 10). Note that $\hat\ung$ is
self-dual if $\ung$ is simply-laced, but otherwise ${\hat\ung}^*$ is a twisted
affine algebra ${\hat\unk}^\sigma$, where $\unk$ is simply-laced and $\sigma$
is a diagram automorphism of $\unk$.

The manifestation of this quantum affine symmetry of interest to us is the
relation, conjectured by physicists, between the so-called \lq fundamental
representations\rq\ of  $U_\epsilon(\hat\ung)$ and the \lq fusings\rq\ of the
classical and quantum particles of the ATFT based on ${\hat\ung}^*$. It is well
known (see [5] and [10], for example) that the masses of the particles in the
theory form the components of the eigenvector with lowest eigenvalue of the
Cartan matrix of $\ung$; in particular, there is a natural one-to-one
correspondence between these particles and the nodes of the Dynkin diagram of
$\ung$. One says that there is a fusing between the particles labelled $i$, $j$
and $k$ if a certain term in the lagrangian of the theory is non-vanishing (see
Section 10). Choose a colouring of the nodes of the Dynkin diagram of $\ung$
black or white in such a way that linked nodes have different colour, and let
$\g$ be the Coxeter element of the Weyl group of $\ung$ obtained by taking the
product of the simp

le reflections associated to the black nodes, followed by those associated to
the white nodes. Let $R_i$ be the $\g$-orbit of the simple root $\alpha_i$ if
$i$ is black, and of $-\alpha_i$ if $i$ is white. Then, Dorey's rule asserts
that there is a non-trivial coupling between the particles labelled $i$, $j$
and $k$ if and only if
$$0\in R_i+R_j+R_k.$$
A little later, it was shown in [20] that (${\roman{D}}$) also gives the fusing
rule for the solitons in the classical theory.

For the theory based on a twisted affine algebra ${\hat\unk}^\sigma$, the
particles are in one-to-one correspondence with the orbits of $\sigma$ on the
nodes of the Dynkin diagram of $\unk$, and a twisted version of (D) is required
to describe their fusings. One defines a \lq twisted Coxeter element\rq\
$\tilde\gamma$ for the pair ($\unk$, $\sigma$), with the property that the
orbits of $\tilde\gamma$ on the set of roots of $\unk$ are in one-to-one
correspondence with the orbits of $\sigma$ on the nodes of the Dynkin diagram
of $\unk$. If $\ung$ is the (non-simply-laced) algebra such that
${\hat\ung}^*\cong{\hat\unk}^\sigma$, these orbits are naturally in one-to-one
correspondence with the nodes of the Dynkin diagram of $\ung$.  Proceeding as
above, one obtains an analogue (TD) of (D), in which the indices $i$, $j$ and
$k$ may be viewed as nodes of the Dynkin diagram of $\ung$, although the
analogues of the $R_i$ are sets of roots of $\unk$. Then the classical fusings
of the ATFT based on ${\hat\ung}^*$, whe

re $\ung$ is non-simply-laced, are given by (TD).

The situation in the quantum theory turns out to be slightly different. This
time, the fusings of the ATFT based on ${\hat\ung}^*$ are given by (D) if
$\ung$ is simply-laced, but by (D) $\cap$ (TD) otherwise (this can be verified
case-by-case using the results in [9], at least when $\ung$ is not of type $E$
or $F$).

Even without this physical motivation, (D) strongly suggests a connection to
representation theory because of its similarity to the condition occurring in
the Parthasarathy--Ranga Rao--Varadarajan (PRV) conjecture [21]. This
conjecture, proved by Kumar [16] and Mathieu [18], asserts that, if $\mu_1$,
$\mu_2$ and $\mu_3$ are dominant weights of $\ung$, $W\mu_1$ the Weyl group
orbit of $\mu_1$, and $W(\mu_1)$ the irreducible $\ung$-module with highest
weight $\mu_1$, etc., then
$$0\in W\mu_1+W\mu_2+W\mu_3 \tag{PRV}$$
implies
$$\text{Hom}_{\ung}(W(\mu_1)\ot W(\mu_2)\ot W(\mu_3),\Bbb C)\ne 0$$
($\Bbb C$ denotes the one-dimensional trivial $\ung$-module). Now, as Braden
[4] pointed out, (D) is equivalent to
$$0\in \Gamma\lambda_i+\Gamma\lambda_j+\Gamma\lambda_k,$$
where $\Gamma$ is the cyclic subgroup of $W$ generated by $\gamma$,  so (D) is
obtained from (PRV) by replacing $W$ by $\Gamma$ (and restricting to
fundamental weights).

The fundamental representations of $U_\epsilon(\hat\ung)$ to which (D) is
related can be characterised as the finite-dimensional irreducible
representations of $U_\epsilon(\hat\ung)$ which contain a fundamental
representation $W_\epsilon(\l_i)$ of $U_\epsilon(\ung)$, and are such that all
other irreducible $U_\epsilon(\ung)$-subrepresentations have highest weight
strictly less than $\l_i$ (see [7] and [8]). There is, in fact, a family of
such representations $V(\l_i,a)$ of $U_\epsilon(\hat\ung)$, depending on a
parameter $a\in\Bbb C^\times$. The representations $V(\l_i,a)$ and $V(\l_i,b)$
are related by twisting by an automorphism of $U_\epsilon(\hat\ung)$ which
fixes $U_\epsilon(\ung)$ and corresponds, at the classical level, to the
automorphism of the loop algebra $\ung[t,t^{-1}]$ which sends $t$ to $at/b$
(the central extension by which $\hat\ung$ is obtained from the loop algebra
plays no role here, since it acts trivially on all the representations of
interest). Now recall that, whether or not $\ung$ is

 simply-laced, the particles of the ATFT based on ${\hat\ung}^*$ are in
one-to-one correspondence with the nodes of the Dynkin diagram of $\ung$. If
$i$, $j$ and $k$ are three such nodes, we would therefore expect a fusing
between the quantum particles labelled $i$, $j$ and $k$ if and only if
$${\roman{Hom}}_{U_\epsilon(\hat{\ung})}(V(\l_i,a)\ot V(\l_j,b)\ot
V(\l_k,c),\Bbb C)\ne 0,\tag{$\ot$}$$
for some $a$, $b$, $c\in \Bbb C^\times$. Thus, ($\ot$) should hold if and only
if $i$, $j$ and $k$ satisfy (D) when $\ung$ is simply-laced, or (D) $\cap$ (TD)
otherwise. This conjecture was first made explicit by MacKay [17].

In this paper, we prove this conjecture when $\ung$ is not of exceptional type.
We also prove an analogous result for Yangians (it was actually in the context
of Yangians that MacKay originally made his conjecture). In fact, in the body
of the paper, we concentrate on the Yangian case, and describe at the end how
to translate the main results from the context of Yangians to that of quantum
affine algebras. As MacKay has emphasized [17], the truth of the conjecture
indicates that there is some beautiful structure in the representation theory
of $U_\epsilon(\hat\ung)$ which is not evident at our present state of
knowledge. It also suggests that it would be interesting to study the
representation theory of twisted quantum affine algebras, but this does not
seem to have been attempted yet.

One approach to the conjecture is through R-matrices. There is a canonical map
$R(a,b)\in{\roman{End}}(V(\l_i,a)\ot V(\l_j,b))$ which is a rational function
of the spectral parameter $a/b$, and is such that $\tau R(a,b)$ commutes with
the action of $U_\epsilon(\hat\ung)$ ($\tau$ denotes the flip of the two
factors in the tensor product). In some cases, explicit formulas for $R(a,b)$
(or rather its Yangian analogue) were given in [7] (and earlier in [19], but
without proper mathematical justification). There is a finite set of values of
$a/b$ for which $R(a,b)$ is well defined, but not invertible, and then its
kernel is a subrepresentation of $V(\l_i,a)\ot V(\l_j,b)$. If one can choose
$a/b$ so that this subrepresentation is fundamental, one deduces that ($\ot$)
holds for some $k$, $c$. To use this method to prove the implication (D) (or
(D) $\cap$ (TD)) $\Rightarrow$ ($\ot$), one would need to compute the R-matrix
associated to every pair of fundamental representations; in addition, one would
have to prove that every fundamental subrepresentation of $V(\l_i,a)\ot
V(\l_j,b)$ arises from

 the R-matrix as above. Because of these difficulties, we employ a different
and simpler method, which makes no use of R-matrices, and which establishes the
reverse implication at the same time.
\vskip12pt\noindent{\it Acknowledgements} We thank Harry Braden and Niall
MacKay for drawing our attention to this problem, and Gustav Delius, Mike
Freeman and Patrick Dorey for several illuminating discussions concerning
ATFTs, and T. A. Springer for help with Coxeter elements, especially in the
twisted case.

\vskip36pt\centerline{\bf 2. Yangians}
\vskip12pt\noindent Let $\ung$ be a finite-dimensional complex semisimple Lie
algebra with Cartan subalgebra $\unh$ and Cartan matrix $A= (a_{ij})_{i,j\in
I}$. Fix coprime positive integers  $(d_i)_{i\in I}$\/ such that the matrix
$(d_ia_{ij})$\/ is symmetric.
Let $R$\/ be the set of roots, $R^+$\/ a set of positive roots, and $R^-=-R^+$.
The roots can be regarded as functions $I\to \Bbb Z$; in particular,
the simple roots $\alpha_i\in R^+$ are given by
$$\alpha_i(j) = a_{ji}, \ \ \ \ \ \  (i,j\in I).$$
Let $Q = \op_{i\in I}\Bbb Z.\alpha_i\subset\unh ^*$\/ be the root lattice, and
set $Q^+ =\sum_{i\in I}\Bbb N.\alpha_i$.

A weight is an arbitrary function $\lambda:I\to\Bbb Z$; denote the set of
weights by $P$, and let
$$P^+ =\{\lambda\in P: \lambda(i)\ge 0 \;\text{for all} \;i\in I\}$$
be the set of dominant weights. Define a partial order $\ge$\/ on $P$\/ by
$$\lambda\ge \mu \;\text{ if and only if}\; \lambda-\mu\in Q^+.$$
Let $\theta$\/ be the unique highest root with respect to $\ge$.

Let $(\ ,\ )$ be the non-degenerate invariant symmetric bilinear form on $\ung$
such that the induced form on $\unh^*$ is given by
$$(\alpha_i,\alpha_j) = d_ia_{ij}.$$
If $\beta\in R$, set $d_\beta =\frac12 (\beta,\beta)$.
Let $W$ be the Weyl group of $\ung$, let $\{s_i\}_{i\in I}$ be the simple
reflections which generate it, and let $w_0$ be the longest element of $W$.
The dual Coxeter number $\check h$ of $\ung$ is
$$\check h=1+2\frac{(\rho,\theta)}{(\theta,\theta)},$$
where $\rho$ is half the sum of the positive roots of $\ung$.

Fix a basis $\{H_i\}_{i\in I}\cup\{X_\alpha^{{}\pm{}}\}_{\alpha\in R^+}$ of
$\ung$ such that, for all $i\in I$, $\alpha,\beta\in R^+$,
$$\align
[H_i,X_\alpha^\pm]=\pm \alpha(i)X_\alpha^\pm,\ \ &
[X_\alpha^+,X_\beta^-]=\delta_{\alpha,\beta}H_\alpha,\\
(H_i,H_j)=d_j^{-1}a_{ij},\ (X_\alpha^+,X_\beta^-)&=\delta_{\alpha,\beta},
\ (X_\alpha^\pm,X_\beta^\pm)=0,
\endalign$$
where $H_\alpha=\sum_in_iH_i$ if $\alpha=\sum_in_i\alpha_i$. Let
$X_i^\pm=X_{\alpha_i}^\pm$.

If $\{I_p\}$ is an orthonormal basis of $\ung$ with respect to $(\ ,\ )$, let
$$\Omega=\sum_pI_p^2$$
be the Casimir element of the universal enveloping algebra $U(\ung)$. We also
denote by $\Omega$ the element
$$\Omega=\sum_pI_p\otimes I_p\in\ung\ot\ung.$$
Let $\k$ be $1/4$ of the value of $\Omega$ acting in the adjoint representation
of $\ung$ (the value of $\k$ is given in Section 3).
\vskip12pt
\proclaim{Definition 2.1} {\rm{([11])}} The Yangian $\y$ is the algebra over
$\Bbb C$ generated by elements
$x$, $J(x)$, for $x\in\frak g$, with the following
defining relations:
$$\align [x,y]\ \ &\text{(in $\y$)}\ =\ [x,y]\ \ \text{(in $\frak
g$)}\,,\tag1\\ J(ax+by)&=aJ(x)+bJ(y)\,,\tag2\\
[x,J(y)]&=J([x,y])\,,\tag3\\
[J(x),J([y,z])]+&[J(z),J([x,y])]+[J(y),J([z,x])]=\\
&\qquad\sum_{p,q,r}([x,I_p]\,,\,
[[y,I_q],[z,I_r]])
\{I_p,I_q,I_r\}\,,\tag4\\
[[J(x),J(y)]\,,\,&[z,J(w)]]+[[J(z),J(w)]\,,\,[x,J(y)]]=\\
& \sum_{p,q,r}([x,I_p]\,,\,[[y,I_q],
[[z,w],I_r]])\{I_p,I_q,J(I_r)\}\,,\tag5
\endalign$$ for all $x$, $y$, $z\in\frak g$, $a$,
$b\in\Bbb C$. Here, for any elements $z_1$, $z_2$, $z_3\in\y$, we set
$$\{z_1,z_2,z_3\}=\frac1{24}\sum_\pi
z_{\pi(1)}z_{\pi(2)}z_{\pi(3)},$$ the sum being over all
permutations $\pi$ of $\{1,2,3\}$.

The Yangian $\y$ has a Hopf algebra structure with counit $\epsilon$,
comultiplication $\Delta$ and antipode $S$ given by
$$\align
\Delta(x)&=x\otimes 1+1\otimes x,\tag6\\
\Delta(J(x))&=J(x)\otimes 1+1\otimes J(x)+\frac12[x\otimes
1\,,\,\Omega],\tag7\\
S(x)&=-x,\ \ S(J(x))=-J(x)+\k x,\tag8\\
\epsilon(x)&=\epsilon(J(x))=0.\tag9\endalign$$
\endproclaim

We shall also need the following presentation of $\y$, given in [12]:
\proclaim{Theorem 2.2} The
Yangian $Y(\frak g)$ is isomorphic to the
associative algebra with generators $X_{i,r}^{{}\pm{}}$,
$H_{i,r}$, $i\in I$, $r\in\Bbb N$, and the following
defining relations:
$$[H_{i,r}\,,\,H_{j,s}]=0,\tag10$$
$$[H_{i,0}\,,\,X_{j,s}^{{}\pm{}}]={}\pm
d_ia_{ij}X_{j,s}^{{}\pm{}},\tag11$$
$$[H_{i,r+1}\,,\,X_{j,s}^{{}\pm{}}]-[H_{i,r}\,,\,X_{j,s+1}^{{}\pm{}}]=
{}\pm\frac12
d_ia_{ij}(H_{i,r}X_{j,s}^{{}\pm{}}+X_{j,s}^{{}\pm{}}H_{i,r}),
\tag12$$
$$[X_{i,r}^+\,,\,X_{j,s}^-]=\delta_{i,j}H_{i,r+s},\tag13$$
$$[X_{i,r+1}^{{}\pm{}}\,,\,X_{j,s}^{{}\pm{}}]-
[X_{i,r}^{{}\pm{}}\,,\,X_{j,s+1}^{{}\pm{}}]={}\pm\frac12
d_ia_{ij}(X_{i,r}^{{}\pm{}}X_{j,s}^{{}\pm{}}
+X_{j,s}^{{}\pm{}}X_{i,r}^{{}\pm{}}),\tag14$$
$$\sum_\pi
[X_{i,r_{\pi(1)}}^{{}\pm{}}\,,\,
[X_{i,r_{\pi(2)}}^{{}\pm{}}\,,\ldots@!,
[X_{i,r_{\pi(m)}}^{{}\pm{}}\,,\,
X_{j,s}^{{}\pm{}}]\cdots]]=0,\tag15$$
for all sequences of non-negative integers $r_1,\ldots@!,r_m$, where
$m=1-a_{ij}$ and the sum is over all permutations $\pi$ of $\{1,\dots,m\}$.

The isomorphism $f$ between the two realizations of
$Y(\frak g)$ is given by

\vbox{
$$\aligned
f(H_i)&=d_i^{-1}H_{i,0},\ \ \
f(J(H_i))=d_i^{-1}H_{i,1}+f(v_i),\\
f(X_i^{{}\pm{}})&=X_{i,0}^{{}\pm{}},\ \ \
f(J(X_i^{{}\pm{}}))=X_{i,1}^{{}\pm{}}+f(w_i^{{}\pm{}}),
\endaligned\tag16$$
where
$$\align
v_i&=\frac14\sum_{\beta\in\Delta^+}
\frac{d_\beta}{d_i}(\beta,\alpha_i)
(X_\beta^+X_\beta^-+X_\beta^-X_\beta^+)-
\frac{d_i}2H_i^2,\\
w_i^{{}\pm{}}&={}\pm\frac14 \!\sum_{\beta\in\Delta^+}\!
d_\beta\left([X_i^{{}\pm{}},X_\beta^{{}\pm{}}]X_\beta^{{}\mp{}}
+X_\beta^{{}\mp{}}[X_i^{{}\pm{}},X_\beta^{{}\pm{}}]\right)
\! -\frac14 \!
d_i(X_i^{{}\pm{}}H_i+H_iX_i^{{}\pm{}}).\!\!\!\qed\endalign$$
}
\endproclaim
\vskip6pt\non{\it Remarks.} 1. The presentation 2.1 of $\y$ shows that there is
a canonical map $\ung\to\y$ (it is known that this map is injective). Thus, any
$\y$-module may be regarded as a $\ung$-module.

2. If $\pi$ is a permutation of $I$ such that
$$H_i\mapsto H_{\pi(i)},\ \ X_i^\pm\mapsto X_{\pi(i)}^\pm$$
defines a Lie algebra automorphism of $\ung$, the assignment
$$H_{i,k}\mapsto H_{\pi(i),k},\ \ X_{i,k}^\pm\mapsto X_{\pi(i),k}^\pm$$
defines a Hopf algebra automorphism of $\y$. We denote both of these
automorphisms simply by $\pi$.
\vskip12pt We shall make use of two further types of automorphism of $\y$.
\proclaim{Proposition 2.3} There is a one-parameter group $\{\tau_a\}_{a\in\Bbb
C}$ of Hopf algebra automorphisms of $\y$ given in terms of the presentation
2.1 by
$$\tau_a(x)=x,\ \ \tau_a(J(x))=J(x)+ax,$$
for $x\in\ung$, and in terms of the presentation 2.2 by
$$\tau_a(H_{i,k})=\sum_{r=0}^k\left({k\atop r}\right)a^{k-r}H_{i,r},\ \ \
\tau_a(X_{i,k}^\pm)=\sum_{r=0}^k\left({k\atop r}\right)a^{k-r}X_{i,r}^\pm.\
\qed$$ \endproclaim

This is Proposition 2.6 in [7].

The second automorphism is an extension of the Cartan involution
$$\varphi_0(H_i)=-H_i,\ \ \ \v_0(X_i^\pm)=X_i^\mp\tag17$$
of $\ung$ to $\y$.
\proclaim{Proposition 2.4} There exists a unique algebra automorphism $\v$ of
$\y$ such that
$$\v(H_{i,k})=(-1)^{k+1}H_{i,k},\ \ \ \v(X_{i,k}^\pm)=(-1)^kX_{i,k}^\mp,$$
for all $i\in I$, $k\in\Bbb N$.
Moreover, $\v$ is a coalgebra anti-automorphism of $\y$.\endproclaim
\demo{Proof} It is easy to check that applying $\v$ to one of the defining
relations in 2.2 gives another of the defining relations. Hence, the assignment
in the statement of the proposition extends uniquely to an algebra homomorphism
$\y\to\y$, and it is obvious that $\v$ is an involution.

Using the isomorphism $f$ in $2.2$, it is clear that $\v\vert_{\ung}=\v_0$ and
that
$$\v(J(H_i))=J(H_i).$$
Hence,
$$\v(J(X_i^\pm))=\mp\frac12\v([X_i^\pm,J(H_i)])=\mp\frac12[X_i^\mp,J(H_i)]
=-J(X_i^\mp).$$
To prove that
$$(\v\ot\v)\circ\Delta=\Delta^{\roman{op}}\circ\v,$$
where $\Delta^{\roman{op}}$ denotes the opposite comultiplication of $\y$, it
suffices to show that both sides agree when applied to a set of generators of
$\y$, such as

\noindent $\{H_i,X_i^\pm,J(H_i),J(X_i^\pm)\}_{i\in I}$. This is now
straightforward, making use of the formula for $\Delta$ in 2.1 and the
observation that $(\v_0\ot\v_0)(\Omega)=\Omega$. $\qed$ \enddemo

We shall also need the following weak version of the Poincar\'e--Birkhoff--Witt
theorem for $\y$.
\proclaim{Proposition 2.5} Let $Y^+$, $Y^-$ and $Y^0$ be the subalgebras of
$\y$ generated by the $X_{i,k}^+$, the $X_{i,k}^-$ and the $H_{i,k}$,
respectively ($i\in I$, $k\in \Bbb N$). Then,
$$\y=Y^-.Y^0.Y^+.\ \ \ \qed$$
\endproclaim

The proof is straightforward.

\vskip36pt\centerline{\bf 3. Finite-dimensional representations}
\vskip 12pt\non
If $W$ is a $\ung$-module and $\lambda\in P$, the weight space
$$W_{\lambda}=\{w\in W|H_i.w=\lambda(i)w\ \ {\text{for all}}\  i\in I\}.$$
If $W_{\lambda}\ne 0$, $\lambda$ is called a weight of $W$, and the set of such
 weights is denoted by $P(W)$.

A non-zero vector $w\in W$ is called a $\ung$-highest weight vector if $w\in
W_{\lambda}$ for some $\lambda\in P(W)$ and $X_i^+.w =0$ for all $i\in I$. Let
$W^+$ be the set of $\ung$-highest weight vectors of $W$, and set $W_\lambda^+
=W^+\cap W_\lambda$. If $W=U(\ung).w$, then $W$ is called a highest weight
$\ung$-module with highest weight $\lambda$. Lowest weight vectors and
$\ung$-modules are defined similarly. For any $\lambda\in P^+$ denote by
$W(\lambda)$ the unique irreducible highest weight $\ung$-module with highest
weight $\lambda$. If $W$ is any finite-dimensional $\ung$-module, we have
$$W\cong\bigoplus_{\lambda\in P^+}W(\lambda)^{\oplus m_\lambda(W)},$$
where the multiplicities $m_{\lambda}(W)$ are given by
$$m_\lambda(W) ={\text{dim}}(W^+\cap W_{\lambda}).$$

We recall that the Casimir operator $\Omega\in U(\ung)$ acts on $W(\lambda)$ by
the scalar $(\lambda+2\rho,\lambda)$. In particular,
$\k=\frac12d_\theta\check h$.

Let $W^*$ be the dual $\ung$-module of $W$, and let $W^{\v}_0$ be the
$\ung$-module obtained by twisting $W$ with the Cartan involution $\v_0$ of
$\ung$.
For $\lambda\in P$, let $\overline{\lambda}=-w_0(\l)$. It is well known that
$$m_{\lambda}(W)=
m_{\overline{\lambda}}(W^*)=m_{\overline{\lambda}}(W^{\v_0}).$$
\vskip12pt
Suppose now that $V$ is a $\uqg$-module. Set
$$V^{++}=\{v\in V^+|X_{i,k}^+.v=0\ \ {\text{for all}}\ \ i\in I, k\in\Bbb
N\},$$
and for any $\lambda\in P^+$, set $V_{\lambda}^{++} =V^{++}\cap V_\lambda$.
Note that, by 2.2, $V^{++}$ is preserved by the action of $Y^0$, and so, if
$V^{++}\ne 0$, it contains a non-zero $Y^0$-eigenvector $v$ (say), so that

$$H_{i,k}.v=d_{i,k}v,$$
for some $d_{i,k}\in\Bbb C$. Such a vector $v$ is called a $\y$-highest weight
vector, $V$ is called $\y$-highest weight if $V=\uqg.v$ for some $\y$-highest
weight vector $v\in V$, and the collection of scalars
$\bold{d}=\{d_{i,k}\}_{i\in I,k\in\Bbb N}$ is called its highest weight. It is
not difficult to show that, for every ${\bold d} =(d_{i,k})_{i\in I,k\in\Bbb
N}$, there is an irreducible $\uqg$-module $V({\bold d})$, unique up to
isomorphism, such that $V({\bold d})$ has highest weight ${\bold d}$. Lowest
weight vectors and modules for $\y$ are defined similarly.

The following theorem of Drinfel'd [12] classifies the finite-dimensional
irreducible $\uqg$-modules.
\proclaim{Theorem 3.1} (i) Every finite-dimensional irreducible $\uqg$-module
is both highest weight and lowest weight.

(ii) If ${\bold d}=(d_{i,k})_{i\in I, k\in\Bbb N}$, the $\uqg$-module $V({\bold
d})$ is finite-dimensional if and only if there exist monic polynomials
$P_i\in\Bbb C[u]$ such that
$$\frac{P_i(u+d_i)}{P_i(u)}=1+\sum_{k=0}^{\infty}d_{i,k}u^{-k-1},\tag18$$
in the sense that the right-hand side is the Laurent expansion of the left-hand
side about $u=\infty$. $\qed$
\endproclaim

If $V$ is a finite-dimensional irreducible $\uqg$-module, we call the
associated $I$-tuple of polynomials $(P_i)_{i\in I}$ the Drinfel'd polynomials
of $V$.

In general, if $V$ is any finite-dimensional $\uqg$-module and $v\in V$ is a
$\y$-highest weight vector, with
$$H_{i,k}.v=d_{i,k}^v v$$
for some $d_{i,k}^v\in\Bbb C$, it follows from 3.1 that there exist monic
polynomials $P_i^v$ such that
$$\frac{P_i^v(u+d_i)}{P_i^v(u)} =1+\sum_{k=0}^{\infty}d_{i,k}^vu^{-k-1}.$$

\proclaim{Proposition 3.2} Let $V_1$, $V_2$ be finite-dimensional
$\uqg$-modules, and let $v_1\in V_1$, $v_2\in V_2$ be $\y$-highest weight
vectors. Then,
$$P_i^{v_1\ot v_2} =P_i^{v_1}P_i^{v_2}.\ \ \qed$$\endproclaim

This is Proposition 2.15 in [7].

The $\y$-modules of interest in this paper are defined as follows.
\proclaim{Definition 3.3} If $i\in I$, $a\in\Bbb C$, then $V_a(\lambda_i)$ is
the finite-dimensional irreducible $\uqg$-module with Drinfel'd polynomials
$$P_j(u)=\cases u-a \ \ \ \ &\text{if $j=i$},\\
1\ \ \ \ &\text{if $j\ne i$}.\endcases$$
We call $V_a(\lambda_i)$ a fundamental $\uqg$-module.\endproclaim

Given a finite-dimensional $\y$-module $V$, we can define the following
associated $\y$-modules:

(i) $V(a)$: this is obtained pulling back $V$ through $\tau_a$;

(ii) $V^\v$:  this is obtained pulling back $V$ through $\varphi$;

(iii) the left dual $^tV$ and right dual $V^t$: these are given by the
following actions of $\y$ on the vector space dual of $V$:
$$\align (y.f)(v)&=f(S(y).v),\ \ \ \ \ y\in\uqg ,f\in {}^tV, v\in V,\\
(y.f)(v)&=f(S^{-1}(y)).v,\ \ \ \ \ y\in\uqg ,f\in V^t, v\in V.\endalign$$
Clearly, if $V$ is irreducible, so are all the representations defined above.

\proclaim{Proposition 3.4.} Let $U$, $V$ and $W$ be finite-dimensional
$\y$-modules, and let $a\in\Bbb C$. Then,
\vskip6pt\non (i) $(U\ot V)^\v\cong V^\v\ot U^\v$;

\non(ii) $V(a)^\v\cong V^\v(-a)$;

\non(iii) ${\roman{Hom}}_{\y}(U,V\ot W)\cong {\roman{Hom}}_{\y}(^t V\ot U, W)$;

\non(iv)  ${\roman{Hom}}_{\y}(U,W\ot V)\cong {\roman{Hom}}_{\y}(U\ot V^t, W)$;

\non(v) $^{tt}V\cong V(-2\k)$,  $V^{tt}\cong V(2\k)$, $^t(V^t)\cong
(^tV)^t\cong V$;

\non(vi) $(V\ot W)^t\cong (W^t\ot V^t)$, $^t(V\ot W)\cong\ {}^tW\ot ^tV$;

\non(vii) $(V(a))^t\cong V^t(a)$, $^t(V(a))\cong (^tV)(a)$. \endproclaim
\demo{Proof} Part (i) follows from the fact that $\v$ is a coalgebra
anti-automorphism of $\uqg$, and part (ii) from the identity
$$\v.\tau_a =\tau_{-a}.\v ,$$
which is proved by checking that the two sides agree when applied to any of the
generators $H_{i,k}$, $X_{i,k}^{{}\pm{}}$. Parts (iii)--(vii) are
straightforward. $\qed$\enddemo

The following result describes the Drinfel'd polynomials of the modules defined
above. If $i\in I$, define $\bar{i}\in I$ by $\l_{\bar i}=\overline{\l_i}$.

\proclaim{Proposition 3.5}  Let $V$ be a finite-dimensional irreducible
$\uqg$-module with Drinfel'd polynomials $P_i$ ($i\in I$), and let $a\in\Bbb
C$.
Then:
\vskip6pt\non
(i) The Drinfel'd polynomials $P_i^a$ of $V(a)$ are given by
$$P_i^a(u)=P_i(u-a).$$

\non(ii) The Drinfel'd polynomials ${}^tP_i$ and $P_i^t$ of $^tV$ and $V^t$,
respectively, are given by
$$^tP_i(u)= P_{\overline{i}}(u+\k),\ \ \
P_i^t(u)=P_{\overline{i}}(u-\k).$$

\non(iii) The Drinfel'd polynomials $P_i^\v$ of ${V^\v}$ are given by
$$P_i^\v(u) = (-1)^{\text{deg}(P_i)}P_{\overline{i}}(\k+d_i-u).$$
\endproclaim
\demo{Proof} Parts (i) and (ii) were proved in [7]. We now prove part (iii).
Let $0\ne v\in V$ be a $\y$-lowest weight vector, and let
$$H_{i,k}.v=\overline{d_{i,k}}v,\ \ \ \  (\overline{d_{i,k}}\in\Bbb C).$$
Then, $v$ is a $\y$-highest weight vector in $V^\v$ and, in $V^\v$, we have, by
2.4,
$$H_{i,k}.v =(-1)^{k+1}\overline{d_{i,k}}v.$$
Hence, the Drinfel'd polynomials $P_i^\v$ of $V^\v$ satisfy
$$\frac{P_i^\v(u+d_i)}{P_i^\v(u)}
=1+\sum_{k=0}^\infty(-1)^{k+1}\overline{d_{i,k}}u^{-k-1}.\tag19$$
On the other hand, by Propositions 3.1 and 3.2 in [7],
$$\frac{P_{\overline{i}}(u-\k)}{P_{\overline{i}}(u+d_i-\k)}
=1+\sum_{k=0}^\infty \overline{d_{i,k}}u^{-k-1}.\tag20$$
The result follows on comparing (19) and (20).\qed\enddemo

\proclaim{Corollary 3.6} Let $i\in I$, $a\in\Bbb C$. Then:
\vskip6pt\non(i) $^t(V_a(\lambda_i))\cong
V_{a-\k}(\lambda_{\overline{i}})$,\hskip1cm
$(V_a(\lambda_i))^t\cong V_{a+\k}(\lambda_{\overline{i}})$;

\non(ii) $(V_a(\lambda_i))^\v\cong V_{\k+d_i-a}(\lambda_{\overline{i}})$.
$\qed$
\endproclaim

We shall also need the following result.
\proclaim{Proposition 3.7} Let $V$ be a finite-dimensional highest weight
$\y$-module. Then, ${V}^\v$ is also a highest weight $\y$-module.
\endproclaim
\demo{Proof} Let $0\ne v\in V_\l$ ($\l\in P^+$) be a $\y$-highest weight
vector. By 2.5, $m_\l(V)=1$ and $m_\mu(V)=0$ unless $\mu\le\l$.
Let $W$ be the $\ung$-submodule of $V$ of type $W(\lambda)$; then $v\in W$. Let
$v^-$ be a lowest weight vector (for $\ung$) in $W$. Then, $v^-$ is a
$\uqg$-highest weight vector in $V^\v$ and
$$\uqg.v^-\supset U(\ung).v^- =W,$$
so $v\in\uqg.v^-$, and hence
$$V^\v=\uqg.v\subset\uqg.v^-.\ \ \ \qed$$
\enddemo

We conclude this section with the following results.
\proclaim{Proposition 3.8} Let $V$ be a finite-dimensional $\uqg$-module. Then,
$V$ is irreducible if and only if $V$ and $^tV$ (resp. $V$ and $V^t$) are both
highest weight $\uqg$-modules.
\endproclaim
\demo{Proof} The \lq only if\rq\  part follows from 3.1 (i). For the converse,
suppose that $V$ and $^tV$ are highest weight (the other case is identical).
Let $v\in V_{\lambda}$ ($\lambda\in P^+$) be a $\y$-highest weight vector. Let
$0\ne W$ be an irreducible $\uqg$-submodule of $V$, and let $\mu$ (say) be the
highest weight of $W$ as a $\ung$-module; thus, $\mu\le\l$. Then, $^tW$ is a
quotient of $^tV$, and these $\ung$-modules have maximal weights $\overline\mu$
and $\overline\lambda$, respectively (cf. the proof of 3.7). Since $^tV$ is a
$\y$-highest weight module, its highest weight vector must map to a non-zero
element of $^tW$. Hence, $\overline\lambda\le\overline\mu$, so $\lambda\le\mu$.
Thus, $\lambda=\mu$ and $W=V$.\qed\enddemo
Along similar lines, we have the following result whose simple proof we omit.
\proclaim{Proposition 3.9} Let $V$ be a finite-dimensional $\uqg$-module, and
assume that, as a $\ung$-module, $V$ has a unique maximal weight $\lambda\in
P^+$. Then, $\uqg.v$ is a proper submodule of $V$ if and only if $V^t$ (resp.
$^tV$) contains a $\y$-highest weight vector of weight strictly less than
$\overline\lambda$. \qed\endproclaim

\vskip36pt\centerline{\bf 4. Dorey's rule}
\vskip12pt\non Let $s_1,s_2,\ldots,s_n$ be the simple reflections in the Weyl
group $W$ of $\ung$ (in some order), and let $\g=s_1s_2\ldots s_n$ be the
associated Coxeter element of $W$. Define positive roots
$$\phi_{i}=s_{n}s_{{n-1}}\ldots s_{i+1}(\alpha_{i}),$$
and let $R_i$ be the $\g$-orbit of $\phi_i$. It is known that the $\phi_i$ are
precisely the positive roots which become negative under the action of $\g$,
and that each $R_i$ contains precisely $h$ roots, where the Coxeter number $h$
is the order of $\g$ (see [15], [22], [23]).

\proclaim{Definition 4.1} If $p\ge 2$, we say that indices
$i_1,i_2,\ldots,i_p\in I$ satisfy condition (${\roman{D}}_{p}$) if and only if
$0\in R_{i_1}+R_{i_2}+\cdots+R_{i_p}$.\endproclaim

Note that the condition (${\roman{D}}_{p}$) appears to depend on a number of
arbitrary choices: we had to pick a Cartan subalgebra $\unh$, a set of positive
roots $R^+$, and an ordering of the set of simple reflections. However, we have

\proclaim{Proposition 4.2} For any $p\ge 2$, the condition (${\roman{D}}_{p}$)
is independent of the choices made.\endproclaim
\demo{Proof} Let $G$ be a (connected, complex) Lie group with Lie algebra
$\ung$. If $\bar{\unh}$ is another Cartan subalgebra, and $\bar{R}^+$ a set of
positive roots with respect to $\bar{\unh}$, there exists $g\in G$ such that
$\bar{\unh}={\roman{Ad}}(g)(\unh)$ and $\bar{R}^+={\roman{Ad}}(g)^*(R^+)$.
Then, the $\bar{\alpha}_i={\roman{Ad}}(g)^*(\alpha_i)$ are the simple roots in
$\bar{R}^+$, and the $\bar{s}_i={\roman{Ad}}(g)\circ s_i\circ
{\roman{Ad}}(g^{-1})$ are the corresponding simple reflections. Using the
Coxeter element $\bar{\gamma}=\bar{s}_1\bar{s}_2\ldots \bar{s}_n$, it is easy
to see that, in an obvious notation, $\bar{R}_i={\roman{Ad}}(g)^*(R_i)$, and it
follows immediately that
$$0\in R_{i_1}+\cdots+R_{i_p}\ \ \ {\roman{iff}}\ \ \ 0\in
\bar{R}_{i_1}+\cdots+\bar{R}_{i_p}.$$

Thus, we may work with a fixed Cartan subalgebra and set of positive roots, and
need only consider the effect of re-ordering the set of simple reflections. It
is well known (see [23], Lemma 2.3) that any such re-ordering can be achieved
by a sequence of moves of the following two types:
\vskip6pt\noindent(i) $s_1s_2\ldots s_{n-1}s_n\mapsto s_ns_1s_2\ldots s_{n-1}$
;

\noindent(ii) $s_1\ldots s_{i-1}s_is_{i+1}s_{i+2}\ldots s_n\mapsto
s_1\ldots s_{i-1}s_{i+1}s_is_{i+2}\ldots s_n$, where $s_is_{i+1}=s_{i+1}s_i$.
\vskip6pt\noindent Thus, it suffices to prove that, if $\bar\gamma$ is the
Coxeter element obtained from $\gamma$ by performing one of these moves, the
condition  (${\roman{D}}_{p}$) obtained by using $\bar\gamma$ is equivelent to
that obtained using $\gamma$. Define $\bar{\phi}_i$ and $\bar{R}_i$ in the
obvious way.

For a move of type (i), it is easy to see that $\bar{\phi}_j=s_n(\phi_j)$ if
$j\ne n$, and $\bar{\phi}_n=s_n\gamma^{-1}(\phi_n)$. Since
$\bar{\gamma}=s_n\gamma s_n$, it follows that $\bar{R}_j=s_n(R_j)$ for all $j$.
It follows as before that the condition (${\roman{D}}_{p}$) is unchanged.

For type (ii), $\bar\gamma=\gamma$ and it is clear that $\bar{\phi}_j=\phi_j$
except possibly when $j=i$ or $i+1$. But
$$\bar{\phi}_i=s_n\ldots s_{i+2}(\alpha_i)=s_n\ldots
s_{i+2}s_{i+1}(\alpha_i)=\phi_i,$$
since $s_{i+1}(\alpha_i)=\alpha_i$, and
$$\bar{\phi}_{i+1}=s_n\ldots s_{i+2}s_i(\alpha_{i+1})=s_n\ldots
s_{i+2}(\alpha_{i+1})=\phi_{i+1},$$
since $s_i(\alpha_{i+1})=\alpha_{i+1}$. Thus, $\bar{R}_j=R_j$ for all $j$.
\qed\enddemo

Despite this result, it is sometimes convenient to make a particular choice of
$\g$, as follows (see [4], [5] and [10], for example). Choose a partition
$$I=I_\circ\amalg I_\bullet\tag21$$
such that
$$a_{ij}=0\ \text{if $i,j\in I_\circ$ or if $i,j\in I_\bullet$}.$$
It is clear that such a partition exists and is unique up to interchanging
$I_\circ$ and $I_\bullet$. Since $s_i$ and $s_j$ commute if $i,j\in I_\circ$ or
if $i,j\in I_\bullet$, the Weyl group elements
$$\g_\circ=\prod_{i\in I_\circ}s_i,\ \ \ \g_\bullet=\prod_{i\in I_\bullet}s_i$$
are well defined. Then we take $\g=\g_\circ\g_\bullet$. Note that
$\g_\circ^2=\g_\bullet^2=1$, so that $\g^{-1}=\g_\bullet\g_\circ$.

With this choice, it is easy to see that
$$\phi_i=\cases \g_\bullet\alpha_i & \text{if $i\in I_\circ$},\\
\alpha_i & \text{if $i\in I_\bullet$}.
\endcases$$
Note that $\g\phi_i=\g_\circ\alpha_i=-\alpha_i$ if $i\in I_\circ$; on the other
hand, if $i\in I_\bullet$, it is clear that $\alpha_i$ occurs with coefficient
$-1$ in the root $\g\phi_i=-\g_\circ\alpha_i$, so $\g\phi_i\in R^-$. It follows
that $\alpha_i\in R_i$ if $i\in I_\bullet$, and $-\alpha_i\in R_i$ if $i\in
I_\circ$.

We observe next that
$$\phi_i=\l_i-\g^{-1}\l_i.$$
Indeed, recalling that $s_i\l_j=\l_j$ if $i\ne j$, and $=\l_i-\alpha_i$ if
$i=j$, the last equation is clear when $i\in I_\bullet$. Similarly, if $i\in
I_\circ$, one has
$$\l_i-\g\l_i=\alpha_i,$$
so that
$$\phi_i=-\g^{-1}(\l_i-\g\l_i)=\l_i-\g^{-1}\l_i.$$
Hence, (${\roman{D}}_{p}$) is equivalent to the condition
$$\g(\g^{r_1}\l_{i_1}+\g^{r_2}\l_{i_2}+\cdots+\g^{r_p}\l_{i_p})=
\g^{r_1}\l_{i_1}+\g^{r_2}\l_{i_2}+\cdots+\g^{r_p}\l_{i_p}$$
for some $r_1,r_2,\ldots,r_p\in\Bbb Z$. Since one is not an eigenvalue of $\g$
on $\unh^*$ (see [15], Lemma 8.1, for example), this last equation is
equivalent to
$$\g^{r_1}\l_{i_1}+\g^{r_2}\l_{i_2}+\cdots+\g^{r_p}\l_{i_p}=0
\tag{${\roman{D}}_{p}'$}$$
for some $r_1,r_2,\ldots,r_p\in\Bbb Z$.

\proclaim{Proposition 4.3.} If $p\ge 2$ and $i_1,i_2,\ldots,i_p\in I$ satisfy
condition (${\roman{D}}_{p}$), then
$${\roman{Hom}}_\ung(W(\l_{i_1})\ot W(\l_{i_2})\ot \cdots\ot W(\l_{i_p}),\Bbb
C)\ne 0.\tag{${\roman{CG}}_{p}$}$$
\endproclaim
\demo{Proof} The condition (${\roman{D}}_{p}'$) is equivalent to
$$\l_{\bar{i}_1}=w_0\g^{r_2-r_1}(\l_{i_2}+\g^{r_3-r_2}\l_{i_3}+\cdots+
\g^{r_p-r_2}\l_{i_p}).$$
By the PRV conjecture, this implies that
$$m_{\l_{\bar{i}_1}}(W(\l_{i_2})\ot\cdots\ot W(\l_{i_p}))\ne 0,$$
or equivalently that
$${\roman{Hom}}_\ung(W(\l_{i_1})\ot W(\l_{i_2})\ot\cdots\ot W(\l_{i_p}),\Bbb
C)\ne 0. \qed$$
\enddemo

This result (and its proof) are due to Braden [4]. A generalisation of it can
also be deduced from the main results of this paper, without using the PRV
conjecture (see the Remark at the end of Section 8).

Proposition 4.6 below shows that the converse of 4.3 is false if $p=3$.  On the
other hand, we have

\proclaim{Proposition 4.4} $(D_2)$ is equivalent to $(CG_2)$.\endproclaim
\demo{Proof} Since
$$\text{Hom}_{\ung}(W(\lambda_i)\ot W(\lambda_j),\Bbb C)\ne 0$$
if and only if $j=\bar{i}$, it suffices to prove that, for all $i\in I$,
$$0\in R_i+R_{\bar{i}}.$$
If the Coxeter number $h$ of $\ung$ is even, it is well known that
$w_0=\gamma^{h/2}$, so
$$\alpha_{\bar{i}}=-\gamma^{h/2}\alpha_i\in -R_i.$$
If $h$ is odd, then $\ung$ is of type $A_n$ (with $n$ even), and the result can
be checked directly by explicitly computing the orbits $R_i$. \qed\enddemo

In the case where $\ung$ is not simply-laced, we shall need a twisted version
of condition (${\roman{D}}_{p}$). For this, we recall that the dual affine Lie
algebra ${\hat\ung}^*$, whose Dynkin diagram is obtained by reversing the
arrows in that of the affine Lie algebra $\hat\ung$, is the twisted affine Lie
algebra associated to a diagram automorphism $\sigma$ of a simply-laced algebra
$\tilde\ung$ (see [14]). Following [22], choose nodes
$\tilde{i}_1,\tilde{i}_2,\ldots,\tilde{i}_n$ of the Dynkin diagram of
$\tilde\ung$, one from each orbit of $\sigma$, and define the {\it twisted
Coxeter element} $\tilde\g$ of $\tilde\ung$ by
$$\tilde\g=\tilde{s}_{\tilde{i}_1}\tilde{s}_{\tilde{i}_2}\ldots
\tilde{s}_{\tilde{i}_n}\sigma$$
(here, and elsewhere in this section, a $\tilde{}\ $ is used to denote objects
associated with $\tilde\ung$). Define roots $\tilde{\phi}_{\tilde{i}_r}$ of
$\tilde\ung$ by
$$\tilde{\phi}_{\tilde{i}_r}=\sigma
\tilde{s}_{\tilde{i}_n}\tilde{s}_{\tilde{i}_{n-1}}\ldots
\tilde{s}_{\tilde{i}_{r+1}}(\tilde\alpha_{\tilde{i}_r}),$$
and let $\tilde{R}_{\tilde{i}_r}$ be the $\tilde\g$-orbit of
$\tilde{\phi}_{\tilde{i}_r}$. Note that there is a natural one-to-one
correspondence between the set $I$ of nodes of the Dynkin diagram of $\ung$ and
the set of orbits of $\sigma$ on the nodes of the Dynkin diagram of
$\tilde\ung$. Thus, if $p\ge 2$ and $i_1,i_2,\ldots,i_p\in I$, the following
condition makes sense:
$$0\in \tilde{R}_{i_1}+\tilde{R}_{i_2}+\cdots+\tilde{R}_{i_p}.
\tag{${\roman{TD}}_{p}$}$$
This is the twisted analogue of condition (${\roman{D}}_{p}$) that we shall
need. As in 4.3, (${\roman{D}}_{p}$) is independent of the choices made in
defining it. It suffices to prove independence of the choice of node from each
$\sigma$-orbit, for a given choice of total ordering of these orbits.
Unfortunately, we have only been able to verify this by a case-by-case check.

As in 4.4, one can check that (${\roman{TD}}_2$) is equivalent to
(${\roman{CG}}_2$) (and hence also to (${\roman{D}}_2$)).

The main purpose of this paper is to study the following conjecture, first made
explicit by MacKay [17] (when $p=3$), but implicit in the work of several
authors on affine Toda field theories (see [5], [10] and [20], for example).

\proclaim{Conjecture 4.5} Let $p\ge 2$ and let $i_1,i_2,\ldots,i_p\in I$. Then,
$${\roman{Hom}}_\y(V_{a_1}(\l_{i_1})\ot V_{a_2}(\l_{i_2})\ot\cdots\ot
V_{a_p}(\l_{i_p}),\Bbb C)\ne 0,\tag23$$
for some $a_1,a_2,\ldots,a_p\in\Bbb C$,
if and only if $i_1,i_2,\ldots,i_p$ satisfy
$$\cases \text{(${\roman{D}}_{p}$) when $\ung$ is simply-laced,}\\
\text{both (${\roman{D}}_{p}$) and (${\roman{TD}}_{p}$) when $\ung$ is not
simply-laced}.
\endcases$$
\endproclaim

It follows immediately from 3.6 and 4.4 that this conjecture is true when
$p=2$. To deal with the $p=3$ case, it is useful to observe that, if $(i,j,k)$
satisfies (23), so does any permutation of $(i,j,k)$ (the same is obviously
true of Dorey's condition)... To see this, note first that, by 3.4 and 3.6 (i),
(23) holds if and only if
$${\roman{Hom}}_\y(V_a(\l_i)\ot V_b(\l_j),V_{c-\k}(\l_{\bar k}))\ne 0,$$
which in turn holds if and only if
$${\roman{Hom}}_\y(V_{c-2\k}(\l_k)\ot V_a(\l_i)\ot V_b(\l_j),\Bbb C)\ne 0.$$
Thus, (23) is preserved by cyclic permutations of $(i,j,k)$. On the other hand,
by 3.6 (ii), (23) is equivalent to
$${\roman{Hom}}_\y(V_{\bar c}(\l_{\bar k})\ot V_{\bar b}(\l_{\bar j})\ot
V_{\bar a}(\l_{\bar i}),\Bbb C)\ne 0,\tag24$$
where $\bar a=\k+d_i-a$, etc. But, it is known that there exists a diagram
automorphism $\pi$ of $\ung$ such that $\pi(i)=\bar i$ for all $i\in I$.
Twisting by the corresponding automorphism of $\y$ shows that (24) is
equivalent to
$${\roman{Hom}}_\y(V_{\bar c}(\l_k)\ot V_{\bar b}(\l_j)\ot
V_{\bar a}(\l_i),\Bbb C)\ne 0.$$
Hence, (23) is also preserved by the permutation $(i,j,k)\mapsto(k,j,i)$. Since
this, together with the cyclic permutations, generates the whole symmetric
group on three letters, (23) is preserved by all permutations of $(i,j,k)$.
It follows that, in proving 4.5, we may always assume that $i$, $j$ and $k$ are
in some fixed order.

We conclude this section by making conditions (${\roman{D_{3}}}$) and
(${\roman{TD_{3}}}$) explicit when $\ung$ is of type $A$, $B$, $C$ or $D$.

\proclaim{Proposition 4.6} Let the nodes of the Dynkin diagram of $\ung$ be
numbered as in [3], and let $1\le i, j\le k\le n$.
\vskip6pt\non (a) Let $\ung$ be of type $A_n$,  Then, $i,j,k$ satisfies
(${\roman{D}}_{3}$) if and only if one of the following conditions holds:

(i) $i+j\le n$, $k=n+1-(i+j)$;

(ii) $i+j>n+1$, $k=2n+2-i-j$.

\vskip6pt\non (b) Let $\ung$ be of type $B_n$ ($n\ge 3$). Then, $i,j,k$
satisfies (${\roman{D}}_{3}$) if and only if one of the following conditions
holds:

(i) $i+j\le n-1$, $k=i+j$;

(ii) $i+j\ge n+1$, $k=2n-i-j$;

(iii) $i<n$, $j=k=n$,

\non and satisfies (${\roman{TD}}_{3}$) if and only if one of the conditions
(i), (iii) or

$\text{(ii)}'$ $i+j\ge n$, $k=2n-1-i-j$

\non holds.
\vskip6pt\non(c) Let $\ung$ be of type $C_n$ ($n\ge 2$). Then, $i,j,k$
satisfies (${\roman{D}}_{3}$) if and only if one of the following conditions
hold:

(i) $i+j\le n$, $k=i+j$;

(ii) $i+j\ge n$, $k=2n-i-j$,

\non and satisfies (${\roman{TD}}_{3}$) if and only if one of the conditions
(i) or

$\text{(ii)}'$ $i+j\ge n+2$, $k=2n+2-i-j$

\non holds.
\vskip6pt\non (d)  Let $\ung$ be of type $D_n$ ($n\ge 4$). Then, $i,j,k$
satisfies (${\roman{D}}_{3}$) if and only if one of the following conditions
holds:

(i) $i+j\le n-2$, $k=i+j$;

(ii) $i+j\ge n$, $k=2n-i-j-2$;

(iii) $i\le n-2$, $n-i$ is even,  $j=k=n-1$ or $j=k=n$;

(iv) $i\le n-2$, $n-i$ is odd,  $j=n-1$, $k=n$.
\qed
\endproclaim

\vskip6pt\non{\it Remarks.}  1. It is interesting to note that, in each of
(a)--(d), case (ii) of condition (${\roman{D}}_3$) can be written $k=h-i-j$,
where $h$ is the Coxeter number of $\ung$, and that in case (b) (resp. (c)),
condition $\text{(ii)}'$ can be written $k=\check{h}-i-j$ (resp.
$2\check{h}-i-j$), where $\check{h}$ is the dual Coxeter number of $\ung$.
(This mysterious factor of 2 in the $C_n$ case is apparently well known to
physicists.)

2. If $\ung$ is of type $D_5$, the triple $2,2,2$ satisfies
(${\roman{CG}}_{3}$) (because $W(\l_2)$ is the adjoint $\ung$-module), but does
not satisfy (${\roman{D}}_{3}$). Thus, the converse of 4.3 is false when $p=3$.
\vskip12pt
The proof of 4.6 is a straightforward, if tedious, computation. We discuss the
example of $\ung=B_4$ to show what is involved. From [14], we see that
$\tilde\ung$ is of type $A_7$ and $\sigma$ is the obvious involution:

\vfill\eject

\hskip2.5cm $\tilde{\ung}$ \hskip5cm $\ung$
\vskip12pt\non We take
$$\g=s_1s_2s_3s_4,\ \
\tilde{\gamma}=\tilde{s}_1\tilde{s}_2\tilde{s}_3\tilde{s}_4\sigma.$$
Then,
$$\phi_1=\alpha_1+\alpha_2+\alpha_3+2\alpha_4,\
\phi_2=\alpha_2+\alpha_3+2\alpha_4,\ \phi_3=\alpha_3+2\alpha_4,\
\phi_4=\alpha_4;$$
$$\tilde\phi_1=\tilde\alpha_4+\tilde\alpha_5+\tilde\alpha_6+\tilde\alpha_7,\
\tilde\phi_2=\tilde\alpha_4+\tilde\alpha_5+\tilde\alpha_6,\
\tilde\phi_3=\tilde\alpha_4+\tilde\alpha_5,\
\tilde\phi_4=\tilde\alpha_4.$$

\topspace{4cm}

\non The orbits are as follows:
$$\align
R_1&=\{\pm \alpha_1,\pm \alpha_2,\pm
\alpha_3,\pm(\alpha_1+\alpha_2+\alpha_3+2\alpha_4)\},\\
R_2&=\{\pm(\alpha_1+\alpha_2),\pm(\alpha_2+\alpha_3),\pm(\alpha_2+\alpha_3+2\alpha_4)\},\\
R_3&=\{\pm(\alpha_3+2\alpha_4),\pm(\alpha_1+\alpha_2+\alpha_3),\pm(\alpha_2+2\alpha_3+2\alpha_4),\pm(\alpha_1+2\alpha_2+2\alpha_3+2\alpha_4)\},\\
R_4&=\{\pm\alpha_4,\pm(\alpha_3+\alpha_4),\pm(\alpha_2+\alpha_3+\alpha_4),
\pm(\alpha_1+\alpha_2+\alpha_3+\alpha_4)\};\endalign$$
$$\align
\tilde{R}_1&=\{\pm\a_1,\pm\a_2,\pm\a_3,\pm\a_6,\pm\a_7,\pm(\a_4+\a_5+\a_6+\a_7),\pm(\a_1+\a_2+\a_3+\a_4+\a_5)\},\\
\tilde{R}_2&=\{\pm(\a_1+\a_2),\pm(\a_2+\a_3),\pm(\a_6+\a_7),\pm(\a_4+\a_5+\a_6),\pm(\a_2+\a_3+\a_4+\a_5),\\
&\phantom{=\{}\pm(\a_3+\a_4+\a_5+\a_6+\a_7),\pm(\a_1+\a_2+\a_3+\a_4+\a_5+\a_6)\},\\
\tilde{R}_3&=\{\pm(\a_4+\a_5),\pm(\a_1+\a_2+\a_3),\pm(\a_3+\a_4+\a_5),\pm(\a_3+\a_4+\a_5+\a_6),\\
&\phantom{=\{}\pm(\a_2+\a_3+\a_4+\a_5+\a_6),\pm(\a_2+\a_3+\a_4+\a_5+\a_6+\a_7),\\
&\phantom{=\{}\pm(\a_1+\a_2+\a_3+\a_4+\a_5+\a_6+\a_7)\},\\
\tilde{R}_4&=\{\pm\a_4,\pm\a_5,\pm(\a_3+\a_4),\pm(\a_5+\a_6),\pm(\a_2+\a_3+\a_4),\pm(\a_5+\a_6+\a_7),\\
&\phantom{=\{}\pm(\a_1+\a_2+\a_3+\a_4)\}.
\endalign$$
By inspection, one sees that $i,j,k$ satisfies (${\roman{D}}_{3}$) or
(${\roman{TD}}_{3}$) exactly in the following cases:
$$\text{(${\roman{D}}_{3}$):}\ \ (1,1,2),\ (1,2,3),\ (1,4,4),\ (2,3,3),\
(2,4,4),\ (3,4,4);$$
$$\text{(${\roman{TD}}_{3}$):}\ \ (1,1,2),\ (1,2,3),\ (1,3,3),\ (1,4,4),\
(2,2,3),\ (2,4,4),\ (3,4,4).$$
These results are in accordance with 4.6.

\vskip36pt
\centerline{\bf 5. Some preliminary lemmas}
\vskip12pt\non
In this section we collect some results which describe the restriction of
$\uqg$-modules to \lq diagram subalgebras\rq\  of $\uqg$.
\proclaim{Definition 5.1} Let $\emptyset\ne J\subseteq I$.
\vskip6pt\non
(i) $\ungj$ is the Lie subalgebra of $\ung$ generated by the $H_i$ and the
$X_i^{{}\pm{}}$ for $i\in J$;

\non
(ii) $Y_J$ is the subalgebra of $\uqg$ generated by the $H_{i,k}$ and the
$X_{i,k}^{{}\pm{}}$, for $i\in J$, $k\in\Bbb N$.

\non
(iii) $Q_J=\sum_{i\in J}\Bbb Z.{\alpha_i}$,\hskip0.5cm$Q_J^+=\sum_{i\in J}\Bbb
N.\alpha_i$.
\endproclaim
It is clear from  2.2 that there is an algebra homomorphism $\uqgj\to Y_J$
which maps $H_{i,k}\mapsto H_{i,k}$ and $X_{i,k}^{{}\pm{}}\mapsto
X_{i,k}^{{}\pm{}}$, for all $i\in J$, $k\in\Bbb N$. In particular, every
$\uqg$-module may be regarded as a $\uqgj$-module. If $V$ is a highest weight
$\uqg$-module with highest weight vector $v$, set
$$V_J=\uqgj.v.$$
 Note that $V_J$ is preserved by the action of $\unh$, since $[\unh,
Y_J]\subseteq Y_J$.
\proclaim{Lemma 5.2} Let $\emptyset\ne J\subseteq I$.
\vskip6pt\non
(i) Let $V$ be a highest weight $\uqg$-module with highest weight $\lambda\in
P^+$ (as a $\ung$-module). Then,
$$V_J=\bigoplus_{\eta\in Q_J^+} V_{\lambda-\eta}.$$
\non
(ii) If $V$ is an irreducible $\uqg$-module, then $V_J$ is an irreducible
$\uqgj$-module.
\non
(iii) If $V$ and $W$ are irreducible $\uqg$-modules with highest weights
$\lambda$ and $\mu$, then,
$$V_J\ot W_J =\bigoplus_{\eta\in Q_J^+} (V\ot W)_{\lambda+\mu-\eta}.\ \ \qed$$
\endproclaim

The proof is straightforward (see Lemma 4.3 in [7] for part (ii)).

The canonical map $\uqgj\to\uqg$ is not a homomorphism of Hopf algebras.
Nevertheless, we have
\proclaim{Lemma 5.3} Let $V$ and $W$ be finite-dimensional irreducible
$\uqg$-modules and let $\emptyset\ne J\subseteq I$. Then, $V_J\ot W_J$ is a
$\uqgj$-submodule of $V\ot W$. \qed
\endproclaim
This is Lemma 2.15 from [7].
The following is a more precise result.
\proclaim{Lemma 5.4} Let $U$, $V$ and $W$ be finite-dimensional irreducible
$\uqg$-modules with highest weights (as $\ung$-modules) $\lambda$, $\mu$ and
$\nu$, respectively, and let $\emptyset\ne J\subseteq I$.
\vskip6pt\non
(i) Assume that $\lambda+\mu-\nu\in Q_J^+$. Then, any non-zero $\uqg$-module
homomorphism $U\ot V\to W$ maps $U_J\ot V_J$ onto $W_J$. In fact, this
restriction defines an injective linear map
$${\roman{Hom}}_{\uqg}(U\ot V, W)\to {\roman{Hom}}_{\uqgj}(U_J\ot V_J,
W_J).\tag25$$
(ii) Assume that $U_J\ot V_J$ is a highest weight $Y(\ung_J)$-module and that
$U\ot V$ has an irreducible quotient $\uqg$-module with highest weight
$\nu<\lambda+\mu$. Then, $\lambda+\mu-\nu\in Q^+\backslash Q_J^+$.
\endproclaim
\demo{Proof} The fact that any $\uqg$-module homomorphism $f:U\ot V\to W$ maps
$U_J\ot V_J$ into $W_J$ follows from 5.2 (i) and (iii). If $f\ne 0$, the image
of $f$ contains a $\uqg$-highest weight vector $w\in W$. By 5.2 (i) and (iii)
again, $w$ is in the image of the restriction of $f$ to $U_J\ot W_J$. By 5.2
(ii), $f$ is surjective, and the linear map (25) is injective.

Part (ii) follows immediately from part (i).
\qed\enddemo
\proclaim{Lemma 5.5} Let $V$ and $W$ be finite-dimensional irreducible
$\uqg$-modules, and let $\emptyset\ne J\subseteq I$. Assume that $V_J\ot W_J$
contains a non-zero $\uqgj$-highest weight vector $u$ which is also an
$\unh$-eigenvector of weight $\nu\in P^+$. Then, $(V\ot W)_\nu$ contains a
$\uqg$-highest weight vector.
\endproclaim
\demo{Proof} Clearly, $\lambda+\mu-\nu\in Q_J^+$. It follows that $u\in (V\ot
W)_{\nu}^{++}$. The result now follows from the discussion preceding 3.1.
\qed\enddemo

\bigpagebreak

\vskip24pt
\centerline{\bf 6. The $A_n$ case}
\vskip 12pt\non
In this section $\ung$ is of type $A_n$ ($n\ge 1$). The Coxeter number $h$ of
$\ung$  is $n+1$. Proposition 4.6 implies that Conjecture 4.5 is a special case
of
\proclaim{Theorem 6.1} Let $1\le i,j,k\le n$, $a,b,c\in\Bbb C$. Then,
$$\roman{Hom}_{\uqg}(V_a(\lambda_i)\ot V_b(\lambda_j), V_c(\lambda_k))\ne
0\tag26$$
if and only if one of the following holds:
\vskip6pt\non
(i) $i+j< n+1$, $k=i+j$, $b-a=\frac12(i+j)$, $c-a=\frac12j$;

\non
(ii) $i+j >n+1$, $k=i+j-n-1$, $b-a=n+1-\frac12(i+j)$,
$c-a=\frac12(n+1-j)$.
\endproclaim
\vskip6pt\non{\it Remark.} It follows from 6.3 (i) below that the space
$$\roman{Hom}_{\uqg}(V_a(\lambda_i)\ot V_b(\lambda_j), V_c(\lambda_k))$$
is one-dimensional when it is non-zero.
\vskip12pt
We shall also prove the following:
\proclaim{Theorem 6.2} Let $1\le i\le j\le n$, $a,b\in\Bbb C$. Then,
$V_a(\lambda_i)\ot V_b(\lambda_j)$ is not a highest weight $\uqg$-module if and
only if $$b-a=\frac12(j-i)+r\ \ \ \ {\text{for some}}\ \ 0<r\le{\roman{min}}
(i,n+1-j).$$
Hence, $V_a(\lambda_i)\ot V_b(\lambda_j)$ is reducible as a $\y$-module if and
only if
$$b-a=\pm\left(\frac12(j-i)+r\right)\ \ \ \ {\text{for some}}\ \
0<r\le{\roman{min}} (i,n+1-j).$$
\endproclaim
\vskip6pt\non{\it Remark.} One can show further that, when $b-a=\frac12(j-i)+r$
for some $0<r\le{\roman{min}}(i,n+1-j)$, the $\y$-module $V_a(\lambda_i)\ot
V_b(\lambda_j)$ has a Jordan--H\"older series of length two:
$$0\to V\to V_a(\lambda_i)\ot V_b(\lambda_j)\to V_{a+\frac12}(\l_{i-r})\ot
V_{a+\frac12(j-i+r)}(\l_{j+r})\to 0,$$
where $V$ is an irreducible $\y$-module such that
$$V\cong\bigoplus_{s=0}^{r-1}W(\l_{i-s}+\l_{j+s})$$
as $\ung$-modules.
\vskip12pt
We begin with the following.
\proclaim{Proposition 6.3} Let $1\le i,j,k\le n$.
\vskip6pt
\noindent (i) We have
$$\roman{Hom}_\ung(W(\lambda_i)\ot W(\lambda_j), W(\lambda_k)) = \cases\Bbb C \
\ {\text{if}} \ \ k=i+j\ \ {\text{or}}\ \ k=i+j-n-1,\\
0 \ \ {\text{otherwise}}.\endcases$$

\noindent(ii)  As $\ung$-modules, we have
$$V_a(\lambda_i)\cong W(\lambda_i).$$

\noindent (iii) Let $a$, $b$, $c\in\Bbb C$. Then,
$${\roman{Hom}}_{\uqg}(V_a(\lambda_i)\ot V_b(\lambda_j), V_c(\lambda_k)) = 0$$
if $k\ne i+j$ or $i+j-n-1$. If $k=i+j$ or $i+j-n-1$, and $a$ and $b$ are fixed,
 the space
$$\roman{Hom}_{\uqg}(V_a(\lambda_i)\ot V_b(\lambda_j), V_c(\lambda_k))\ne 0$$
for at most one value of $c$, in which case it is one-dimensional.
\endproclaim
\demo{Proof} Part (i) is easy, part (ii) is well known (see [11] and [8]), and
part (iii) is immediate from parts (i) and (ii).
\qed\enddemo
\demo{Proof of 6.1} By induction on $n$. The case $n=1$ is proved in [6].
Twisting by $\v$ and using 3.5, we see that
$$ \text{Hom}_{\y}(V_a(\lambda_i)\ot V_b(\lambda_j), V_c(\lambda_k))\ne 0$$
if and only if
$$\text{Hom}_{\y}(V_{\frac12(n+3)-b}(\lambda_{n+1-j})\ot
V_{\frac12(n+3)-a}(\lambda_{n+1-i}), V_{\frac12(n+3)-c}(\lambda_{n+1-k}))\ne
0.$$ Hence  it suffices to prove the theorem when $i+j <n+1$.

Assume that
$$\text{Hom}_{\y}(V_a(\l_i)\ot V_b(\l_j),V_c(\l_k))\ne 0.\tag27$$
Since $i+j<n$,  6.3 implies that $k=i+j$.  Noting that
$$\lambda_i+\lambda_j-\lambda_{i+j}\in Q_{J}^+,$$
where $J=\{1,2,\ldots ,i+j-1\}$, 5.4 gives
$$\text{Hom}_{Y(\ung_{J})}(V_a(\lambda_i)\ot V_b(\lambda_j), \Bbb C)\ne 0,$$
whence $b-a=\frac12(i+j)$ by 3.6.

The value of $c$ can be computed as follows. Using 3.5 and 3.6, (27) implies
that
$$\text{Hom}_{\y}(V_{a+\frac12(i+j)}(\lambda_j),
V_{a+\frac12(n+1)}(\lambda_{n+1-i})\ot V_{c}(\lambda_{i+j}))\ne 0$$
and hence, by taking left duals, that
$$\text{Hom}_{\y}(V_{c-\frac12(n+1)}(\lambda_{n+1-i-j})\ot V_a(\lambda_i),
V_{a-\frac12(n+1-i-j)}(\lambda_{n+1-j}))\ne 0.$$
The first part of the proof now shows that
$$\align
a-\left(c-\frac12(n+1)\right)&=\frac12(i+n+1-i-j),\\
\noalign{i.e.} c-a&=\frac12j.\endalign$$

For the \lq if\rq\ part, suppose that
 $k=i+j$, $b-a=\frac12(i+j)$ and $c-a=\frac12j$. Using 5.4 with $J=\{n-i-j-2,
\ldots ,n\}$ and 5.5, we see that
$$(V_{b-\frac12(n+1)}(\lambda_{n+1-j})\ot
V_{a-\frac12(n+1)}(\lambda_{n+1-i}))^{++}_{\lambda_{n+1-i-j}}\ne 0.$$
Since there is no non-zero dominant weight strictly less than
$\lambda_{n+1-i-j}$, it follows that for some $c'\in\Bbb C$,
$$\text{Hom}_{\y}(V_{c'-\frac12(n+1)}(\lambda_{n+1-i-j}),
V_{b-\frac12(n+1)}(\lambda_{n+1-j})\ot V_{a-\frac12(n+1)}(\lambda_{n+1}))\ne
0.$$
Applying 3.5 and 3.6 shows that
$$\text{Hom}_{\y}(V_a(\lambda_i)\ot V_b(\lambda_j), V_{c'}(\lambda_k))\ne 0.$$
But then, by 6.3 (iii), $c'$ is uniquely determined, and by the \lq only if\rq\
 part, $c' =c$.

The proof of 6.1 is now complete.
\qed\enddemo

\demo{Proof of 6.2} By induction on $n$. If $n=1$, the result is contained in
[6]. Assuming the result is known when $\ung$ is of type $A_m$ for $m<n$, we
prove it when $\ung$ is of type $A_n$ by induction on ${\text{min}}(i,n+1-j)$.
If $i=1$ or $j=n$, the result follows from 6.1 and 6.3, since
$$\align W(\lambda_1)\ot W(\lambda_j)&\cong W(\lambda_1+\lambda_j)\oplus
W(\lambda_{j+1}),\\
W(\lambda_n)\ot W(\lambda_j)&\cong W(\lambda_n+\lambda_j)\oplus
W(\lambda_{j-1}).\endalign$$

Assume now that ${\text{min}}(i,n+1-j)>1$. To prove the \lq only if\rq\ part of
6.2, consider the case $i+j< n+1$ (resp. the case $i+j>n+1$).

Since $V_a(\lambda_i)\ot V_b(\lambda_j)$ is not a highest weight $\uqg$-module,
there exists an irreducible $\uqg$-module $V$ with $\ung$-highest weight
$\lambda =\lambda_i+\lambda_j-\eta$, for some $0\ne \eta\in Q^+$, such that
$$\text{Hom}_{\y}(V_a(\lambda_i)\ot V_b(\lambda_j), V)\ne 0.$$
By 3.5, 3.6 and 6.1, we have
$$\text{Hom}_{\y}(V_{a+\frac12}(\lambda_{i-1})\ot V_b(\lambda_j),
V_{a+\frac12(n-i)}(\lambda_n)\ot V)\ne 0,$$
(resp. $$\text{Hom}_{\y}(V_{a}(\lambda_{i})\ot V_{b-\frac12}(\lambda_{j+1}),
V\ot V_{b-\frac12(j+1)}(\lambda_1))\ne 0).$$
If $b-a\ne\frac12(j-i)+r$ for any $1<r\le i$, then by the induction hypothesis
on ${\text{min}}(i,n+1-j)$, $V_{a+\frac12}(\lambda_{i-1})\ot V_b(\lambda_j)$
(resp. $V_a(\lambda_i)\ot V_{b-\frac12}(\lambda_{j+1})$) is a highest weight
$\uqg$-module, so
$$\lambda_{i-1}+\lambda_j\le \lambda_n+\lambda_i+\lambda_j-\eta ,$$
i.e. $$\eta\le \alpha_i+\cdots +\alpha_n,$$
(resp. $$\lambda_i+\lambda_{j+1}\le\lambda_1+\lambda_i+\lambda_j-\eta,$$
i.e. $$\eta\le\alpha_1+\cdots+\alpha_j{\roman{)}}.$$
This, together with the requirement that $\lambda\in P^+$, forces
$\eta=\alpha_i+\cdots +\alpha_j$, so $\lambda=\lambda_{i-1}+\lambda_{j+1}$.
Noting that $\lambda_i+\lambda_j-\lambda_{i-1}-\lambda_{j+1}\in Q_J^+$, where
$J=\{i,i+1,\ldots ,j\}$, it follows from 5.4 that
$$\text{Hom}_{\uqgj}(V_a(\lambda_1)\ot V_b(\lambda_{j-i+1}), \Bbb C)\ne 0.$$
By 3.6, we see that $b-a=\frac12(j-i)+1$, as required.

We now prove the \lq if\rq\ part of 6.2, assuming it when $\ung$ is of type
$A_m$ for $m<n$, and for smaller values of ${\text{min}}(i,n-j+1)$ when $\ung$
is of type $A_n$. We consider three cases.

Suppose first that $i+j<n+1$ (resp. $i+j>n+1$). Let
$$b-a=\frac12(j-i)+r \ \ {\text{for some}}\ \ 0<r\le i,\tag28$$
and assume for a contradiction that $V_a(\lambda_i)\ot V_b(\lambda_j)$
is a highest weight $\uqg$-module. Let $J'=\{1,2,\ldots ,n-1\}$ (resp.
$J'=\{2,\ldots ,n\}$). Since $V_a(\lambda_i)\ot V_b(\lambda_j)$ is assumed to
be $\y$-highest weight,
$$V_a(\lambda_i)_{J'}\ot V_b(\lambda_j)_{J'}\subset \uqg.(v_i\ot v_j),$$
where $v_i$ and $v_j$ are $\y$-highest weight vectors in $V_a(\l_i)$ and
$V_b(\l_j)$. But then 5.2 implies that
$$V_a(\lambda_i)_{J'}\ot V_b(\lambda_j)_{J'}\subset Y(\ung_{J'}).(v_i\ot
v_j),$$
and hence that $V_a(\lambda_i)_{J'}\ot V_b(\lambda_j)_{J'}$ is
$Y(\ung_{J'})$-highest weight. By the induction hypothesis on $n$, $b-a$ cannot
take any of the values in (28). This is the desired contradiction.

If  $i+j=n+1$, the argument used above fails when $b-a=\frac12(n+1)$, since in
that case $V_a(\lambda_i)_{J'}\ot V_b(\lambda_j)_{J'}$ is
$Y(\ung_{J'})$-highest weight. But for this value of $b-a$, the contradiction
is immediate from 3.6.

We have now completely proved Theorem 6.2, except for the final statement,
which follows immediately from 3.6 and 3.8.\qed\enddemo

\vskip24pt\centerline{\bf 7. The $D_n$ case}.
\vskip12pt\non
In this section $\ung$ is of type $D_n$, ($n\ge 4$). The Coxeter number $h$ is
$2n-2$. Conjecture 4.5 is a special case of
\proclaim{Theorem 7.1} Let $1\le i\le j\le k\le n$, $a,b,c\in\Bbb C$. Then,
$${\roman{Hom}}_{\y}(V_a(\lambda_i)\ot V_b(\lambda_j), V_c(\lambda_k))\ne 0$$
if and only if one of the following holds:
\vskip6pt\non
(i) $i+j\le n-2$, $k=i+j$, $b-a=\frac12(i+j)$, $c-a=\frac12 j$;

\non
(ii) $i+j\ge n$, $j\le n-2$, $k=2n-i-j-2$, $b-a=\frac12(i+j)$, $c-a=\frac12j$;

\non
(iii) $i\le n-2$, $j=n-1$, $b-a=\frac12(n+i-1)$, $c-a=\frac12(n-i-1)$,
$$ k=\cases\overline{n-1}\ \ &{\text{if}}\ \ \ \ n-i\ \ {\text{is even}},\\
\overline{n}\ \ &{\text{if}}\ \ \ \  n-i \ \ {\text{is odd}};\endcases$$
(iv) $i\le n-2$, $j=n$, $b-a=\frac12(n+i-1)$, $c-a=\frac12(n-i-1)$,
$$ k=\cases\overline{n}\ \ &{\text{if}}\ \ n-i\ \ {\text{is even}},\\
\overline{n-1}\ \ &{\text{if}}\ n-i \ \ {\text{is odd}}.\endcases$$
Moreover, the space of homomorphisms in (26) is one-dimensional when it is
non-zero.
\endproclaim

We shall also prove
\proclaim{Theorem 7.2} Let $1\le i\le j\le n$. Then, $V_a(\lambda_i)\ot
V_b(\lambda_j)$ is not a highest weight $\uqg$-module if and only if one of the
following holds:
\vskip6pt\non
(i) $j\le n-2$,
$$b-a=\cases\frac12(j-i)+r\ \ &\text{for some $0<r\le{\roman{min}}(i,n-j)$,
or}\\
n-1-r-\frac12(j-i) \ \ &\text{for some $0\le r <
{\roman{min}}(i,n-j)$;}\endcases$$

\non
(ii) $i\le n-2$, $j=n-1$ or $n$,
$$b-a=\frac12(n-1-i)+r \ \ \text{for some $0<r\le i$};$$
(iii) $i=j=n-1$ or $n$,
$$b-a=n-r-1 \ \ \ \ \text{for some $0\le r\le n-2$ with $ n-r$ even;}$$
(iv) $i=n-1$, $j=n$,
$$b-a=n-r-1 \ \ \ \ \text{for some $0\le r\le n-2$ with $n-r$ odd}.$$
Hence, $V_a(\lambda_i)\ot V_b(\lambda_j)$ is reducible as a $\y$-module if and
only if $\pm(b-a)$ takes one of the above values.
\endproclaim

We first recall from [7], Theorem 6.2, the $\ung$-module structure of the
fundamental $\uqg$-modules.
\proclaim{Proposition 7.3} Let $a\in\Bbb C$. Then as a $\ung$-module,
$$V_a(\lambda_i)\cong\cases \bigoplus_{k=0}^{[\frac{i}{2}]}W(\lambda_{i-2k})\ \
\ \ &\text{if $i\le n-2$},\\
W(\lambda_i)\ \ \ \ &\text{if $i=n-1$ or $n$,}\endcases$$
where $\lambda_0=0$. \qed
\endproclaim
We first prove
\proclaim{Proposition 7.4} Let $1\le j\le n-2$ and let $v_1$ and $v_j$ be
$\y$-highest weight vectors in $V(\lambda_1)$ and $V(\lambda_j)$, respectively.
\vskip6pt\non
(i) $V_a(\lambda_1)\ot V_b(\lambda_j)$ is not a $\y$-highest weight module if
and only if $b-a=\frac12(j+1)$ or $n-\frac12(j+1)$.
\vskip6pt\non
(ii) If $b-a =\frac12(j+1)$ or $n-\frac12(j+1)$), then $V_a(\lambda_1)\ot
V_b(\lambda_j)$ has a Jordan--H\"older series of length two, namely
$$0\to\ \uqg.(v_1\ot v_j)\to   V_a(\lambda_1)\ot V_b(\lambda_j)\to
V_{a+\frac12j}(\lambda_{j+1})\to 0,$$
or
$$0\to\ \uqg.(v_1\ot v_j)\to   V_a(\lambda_1)\ot V_b(\lambda_j)\to
V_{a+n-1-\frac12j}(\lambda_{j-1})\to 0),$$
respectively (if $j=n-2$, the first short exact sequence should be replaced by
$$0\to\ \uqg.(v_1\ot v_j)\to   V_a(\lambda_1)\ot V_b(\lambda_j)\to
V_{a+\frac12(n-2)}(\lambda_{n-1})\ot V_{a+\frac12(n-2)}(\l_n)\to 0).$$
\endproclaim
\demo{Proof} Let $b-a=\frac12(j+1)$ and $J=\{1,2,\ldots ,n-2\}$. By 6.1, the
$\uqg$-submodule $V_b(\lambda_j)_J\ot V_a(\lambda_1)_J$ of $V_b(\lambda_j)\ot
V_a(\lambda_1)$ has a $\uqgj$-highest weight vector of weight
$\lambda_{j+1}=\lambda_1+\lambda_j-\alpha_1-\cdots -\alpha_j$ for $\ung$.
Hence, by 5.5, $V_b(\lambda_j)\ot V_a(\lambda_1)$ has a $\uqg$-highest weight
vector of weight $\lambda_{j+1}$. But then, by 3.14, $V_a(\lambda_1)\ot
V_b(\lambda_j)$ cannot be $\y$-highest weight.

For the converse, assume that $V_a(\lambda_1)\ot V_b(\lambda_j)$ is not
$\y$-highest weight, let $M=V_a(\lambda_1)\ot V_b(\lambda_j)/\y.(v_1\ot v_j)$,
let $N$ be an irreducible quotient of $M$, and let $\lambda\in P^+$ be the
maximal weight of $N$ as a $\ung$-module.
The dominant weights $\lambda<\lambda_1+\lambda_j$ are of two types:
\vskip6pt\non
(i) $j\ge 3$, $\lambda=\lambda_1+\lambda_{j-2k}$, $0<k\le[j/2]$, and

\non(ii) $\lambda =\lambda_{j+1-2k}$, $0\le k\le[(j+1)/2],$
\vskip6pt\non
with the understanding that $\lambda_0=0$. In case (i),
$${\text{Hom}}_\y(V_a(\lambda_1)\ot V_b(\lambda_j), N)\ne 0,$$
which implies that
$${\text{Hom}}_\y(V_b(\lambda_j), V_{a+n-1}(\lambda_1)\ot N)\ne 0,$$
and hence that $\lambda_j\le 2\lambda_1+\lambda_{j-2k}$. This implies that
$k=1$,  i.e.
$\lambda=\lambda_1+\lambda_{j-2}$. But then $\lambda_1+\lambda_j-\lambda\in
Q_{J'}^+$, where $J'=\{j-1,j,\ldots ,n\}$. Hence, 5.5 implies that
$${\text{Hom}}_{Y(\ung_{J'})}(V_b(\lambda_2),\Bbb C)\ne 0,$$
which is absurd.
Thus, case (ii) must hold. As above, one sees that $k=0$ or 1 (and $k=1$ if
$j=n-2$), so we must have either
\vskip6pt\non
\noindent (iia) ${\text{Hom}}_{\y}(V_a(\lambda_1)\ot V_b(\lambda_j),
V_c(\lambda_{j+1}))\ne 0$ (if $j<n-2$), or

\noindent(iib) ${\text{Hom}}_{\y}(V_a(\lambda_1)\ot V_b(\lambda_j),
V_c(\lambda_{j-1}))\ne 0$,
\vskip6pt\non for some $c\in\Bbb C$.

In case (iia), note that $\lambda_1+\lambda_j-\lambda_{j+1}=\alpha_1+\cdots
+\alpha_j\in Q_{J''}^+$, where $J''=\{1,2,\ldots ,n-2\}$. Hence,
$${\text{Hom}}_{Y(\ung_{J''})}(V_a(\lambda_1)_{J''}\ot V_b(\lambda_j)_{J''},
V_c(\lambda_{j+1})_{J''})\ne 0,$$
which gives $b-a=\frac12(j+1)$ and $c-a=\frac12j$, by 6.1.

In case (iib), we get
$${\text{Hom}}_\y(V_b(\lambda_j), V_{a+n-1}(\lambda_1)\ot
V_c(\lambda_{j-1}))\ne 0,$$
and hence, taking left duals,
$${\text{Hom}}_\y(V_{c-n+1}(\lambda_{j-1})\ot V_a(\lambda_1),
V_{b-n+1}(\lambda_j))\ne 0.$$
Finally, twisting with $\v$ gives
$${\text{Hom}}_\y(V_{n-a}(\lambda_1)\ot V_{2n-1-c}(\lambda_{j-1}),
V_{2n-1-b}(\lambda_j))\ne 0.$$
We are now in the situation of (iia). Hence,
$$2n-1-c-(n-a)=\frac12 j\ \ \text{and}\ \ \ 2n-1-b-(n-a)=\frac12(j-1),$$
i.e.
$$b-a=n-\frac12(j+1)\ \ \text{and}\ \ \ c-a=n-1-\frac12 j.$$

We have now proved (i). In fact, the preceding argument shows that, if $V$ is
an  irreducible quotient of $V_a(\lambda_1)\ot V_b(\lambda_j)$ with highest
weight different from $\lambda_1+\lambda_j$, then either
$$\aligned
b-a=\frac12(i+j)\ \ &\text{and}\ \ V\cong V_{a+\frac12 j}(\lambda_{j+1}),\
\text{or}\\
b-a=n-\frac12(j+1)\ \ &\text{and}\ \
V\cong V_{a+n-1-\frac12j}(\lambda_{j-1}).\endaligned\tag29$$

We prove part (ii) of 7.4 when $b-a=\frac12(j+1)$; the other case is similar.
First, if $N$ is any $\y$-submodule (or quotient module) of $V_a(\lambda_1)\ot
V_b(\lambda_j)$, then, since ${\roman{Hom}}_\y(V_{a-n+1}(\lambda_1)\ot N,
V_b(\lambda_j))\ne 0$,
(resp. ${\roman{Hom}}_\y(^tN\ot V_a(\lambda_1), V_{b-n+1}(\lambda_j))\ne 0$),
we see by using 7.3 that
$$m_0(N)\ne 0 \ \ \text{if $j$ is odd, and $m_1(N)\ne 0$ if $j$ is even}.$$

If $L =\uqg.(v_1\ot v_j)$ is  reducible for $\y$, let $L'\subset L$ be an
irreducible $\uqg$-submodule. Then, $L'\cong V_e(\lambda_{j-1})$ for some
$e\in\Bbb C$, and we get
$${\text{Hom}}_\y(V_e(\lambda_{j-1}), V_a(\lambda_1)\ot V_b(\lambda_j))\ne 0.$$
But this is impossible when $b-a=\frac12(j+1)$, by (29).

Let $M$  be the quotient $V_a(\lambda_1)\ot V_b(\lambda_j)/L$. Then,
$M=\uqg.M_{\lambda_{j+1}}$, since otherwise $M$ would have an irreducible
quotient which would have to be of highest weight  $\lambda_{j-1}$, and we have
seen above that this is impossible for this value of $b-a$. Since $M$ is
non-zero, this shows that $M_{\lambda_{j+1}}\ne 0$. On the other hand,  since
$M_{\lambda_{j+1}}\subseteq M^+$, we have
$${\text{dim}}(M_{\lambda_{j+1}})\le m_{\lambda_{j+1}}(V_a(\lambda_1)\ot
V_b(\lambda_j)).$$
This multiplicity is one, and so $M_{\lambda_{j+1}}$ is one-dimensional.
Thus, $M$ is a highest weight $\y$-module with $\ung$-highest weight
$\lambda_{j+1}$. If $M$ is not irreducible for $\y$, it contains an irreducible
$\y$-submodule, which must be of the form $V_d(\lambda_{j+1-2k})$ for some
$1\le k\le[(j+1)/2]$, $d\in\Bbb C$. By 7.3, this means that $m_{\lambda_{1}}(M)
=2$ if $j$ is even, and $m_{0}(M)=2$ if $j$ is odd. But this would mean that
$m_0(L) =0$ or $m_{\l_1}(L) =0$, and we have seen that this is impossible.
\qed\enddemo

To prove 7.1, we need
\proclaim{Proposition 7.5} Let $1\le i\le n-2$, $a\in\Bbb C$.
\vskip6pt\non
(i) If $n-i$ is even,
$$\align
{\roman{Hom}}_\y(V_a(\lambda_{n})\ot V_b(\lambda_{n}), V_c(\lambda_i))\ne 0 \ \
& {\text{iff}}\ \  b-a=n-i-1, \ c-a=\frac12(n-i-1),\\
{\roman{Hom}}_\y(V_a(\lambda_{n-1})\ot V_b(\lambda_{n-1}), V_c(\lambda_i))\ne 0
\ \  &{\text{iff}}\ \  b-a=n-i-1,\  c-a=\frac12(n-i-1).
\endalign$$

\non(ii) If $n-i$ is odd,
$$\align
{\roman{Hom}}_\y(V_a(\lambda_{n-1})\ot V_b(\lambda_n), V_c(\lambda_i))\ne 0 \ \
& {\text{iff}}\ \ b-a=n-i-1,\ \ c-a=\frac12(n-i-1),\\
{\roman{Hom}}_\y(V_a(\lambda_{n})\ot V_b(\lambda_{n-1}), V_c(\lambda_i))\ne 0 &
\ \ {\text{iff}}\ \  b-a=n-i-1, \ \ c-a=\frac12(n-i-1).
\endalign $$
\endproclaim
\demo{Proof} We prove the first statement in part (i); the proofs in the other
cases are similar. In [7], Proposition 6.2, we established that, if
$b-a=n-i-1$, there exist $c$, $c'\in\Bbb C$ such that
$${\text{Hom}}_\y(V_a(\lambda_{n})\ot V_b(\lambda_n), V_c(\lambda_i))\ne 0\
\text{and}\  {\text{Hom}}_\y(V_a(\lambda_{n-1})\ot V_b(\lambda_{n-1}),
V_{c'}(\lambda_i))\ne 0.$$
To see that $c=c' =a+\frac12(n-i-1)$, notice that since
$m_{\lambda_i}(V(\lambda_n)\ot V(\lambda_n)) =1$, the values of $c$ and $c'$
are uniquely determined by $a$ and $b$. But now, twisting by $\v$ and applying
$\tau_{a+b}$ gives
$${\roman{Hom}}_{\y}(V_a(\lambda_{n})\ot V_b(\lambda_n),
V_{a+b-c}(\lambda_i))\ne 0$$
and
$${\roman{Hom}}_{\y}(V_a(\lambda_{n-1})\ot V_b(\lambda_{n-1}),
V_{a+b-c'}(\lambda_i))\ne 0,$$
and hence
$$a+b-c=c\ \ \ \text{and}\ \ \ a+b-c'=c'.$$

Conversely, suppose that ${\roman{Hom}}_{\uqg}(V_a(\lambda_n)\ot
V_b(\lambda_n), V_c(\lambda_i))\ne 0$. We prove by induction on $n$ that $b-a$
and $c-a$ have the stated values. If $n=4$, the result follows from 7.4 (i) by
using a diagram automorphism of order three of $\y$, so the induction begins.
Assume the result when $\ung$ is of type $D_m$ with $m<n$. Now,
$2\lambda_n-\lambda_i\in Q_J^+$,  where $J=\{2,3,\ldots ,n\}$, so by the
induction hypothesis on $n$, we get
$$b-a=n-1-(i-1)-1=n-i-1.$$
The value of $c-a$ is determined as before.
\qed\enddemo

\demo{Proof of Theorem 7.1}  We only have to prove the theorem in cases (i) and
(ii), since 7.5 establishes cases (iii) and (iv).

 The \lq only if\rq\ part is proved by by induction on $n$. The induction
actually begins at $n=3$, when $\ung$ is of type $A_3$, and the result in that
case is contained in 6.1. Assume now that $n\ge 4$ and that the result is known
when $\ung$ is of type $D_m$ for $m<n$.

Suppose then that
$${\text{Hom}}_\y(V_a(\lambda_i)\ot V_b(\lambda_j), V_c(\lambda_k))\ne
0.\tag30$$
This implies by Proposition 7.4 that
$${\text{Hom}}_\y(V_{a-\frac12}(\lambda_{i-1})\ot
V_{a+\frac12(i-1)}(\lambda_1)\ot V_b(\lambda_j), V_c(\lambda_k))\ne 0,$$
and hence
$${\text{Hom}}_\y(V_{a+\frac12(i-1)}(\lambda_1)\ot V_b(\lambda_j),
V_{a+n-\frac32}(\lambda_{i-1})\ot V_c(\lambda_k))\ne 0.$$
Let $F$ be a non--zero element in
${\text{Hom}}_\y(V_{a+\frac12(i-1)}(\lambda_1)\ot V_b(\lambda_j),
V_{a+n-\frac32}(\lambda_{i-1})\ot V_c(\lambda_k))$, and let $v_1$ and $v_j$ be
$\y$-highest weight vectors in $V_{a+\frac12(i-1)}(\lambda_1)$ and
$V_b(\lambda_j)$, respectively. We first  prove that one of the following must
hold:

\vskip6pt
\noindent ($\alpha$) $b-a=\frac12(i+j)$ and $F(v_1\ot v_j)=0$;
\vskip6pt
\noindent ($\beta$)  $b-a=n-1-\frac12(j-i)$ and $F(v_1\ot v_j) =0$;
\vskip6pt
\noindent ($\gamma$) $F(v_1\ot v_j)\ne 0$, $i+j\le n$, $k=i+j-2$,
$b-a=\frac12(i+j)-1$, $c-a=\frac12j$;
\vskip6pt
\noindent($\delta$) $F(v_1\ot v_j)\ne 0$, $i+j\ge n+2$, $ k=2n-i-j$,
$b-a=\frac12(i+j)-1$, $c-a=\frac12j$.
\vskip6pt

If $F(v_1\ot v_j) =0$, then,  by 7.4, we see that either $(\alpha)$ or
$(\beta)$
must hold. On the other hand, if $F(v_1\ot v_j)\ne 0$, then, using the fact
that
$\lambda_{i-1}+\lambda_k-\lambda_1-\lambda_j\in Q_J^+$, where $J=\{2,3,\ldots
,n\}$, we see from 5.5 that, if $i>2$,
$${\text{Hom}}_{\uqgj}(V_{a+\frac12}(\lambda_{i-2})\ot V_b(\lambda_{j-1}),
V_c(\lambda_{k-1}))\ne 0.$$
The induction hypothesis on $n$ now shows that either ($\gamma$) or ($\delta$)
must hold. If $i=1$, the same conclusion follows from 7.4. Finally, if $i=2$,
we get $b=c$ and $j=k$, so
$$F:V_{a+\frac12}(\l_1)\ot V_b(\l_j)\to V_{a+n-\frac32}(\l_1)\ot V_b(\l_j).$$
Since $a+\frac12\ne a+n-\frac32$, 3.2 implies that $F(v_1\ot v_j)=0$,
contradicting our assumption. This completes the proof that one of
($\alpha$)--($\delta$) must hold.

Next,  we prove that (30) implies that $(\alpha)$ must hold. Observe that if we
twist by  $\v$ and apply $\tau_{a+b-n}$, then (30) implies that
$${\text{Hom}}_\y(V_a(\lambda_j)\ot V_b(\lambda_i), V_{a+b-c}(\lambda_k))\ne
0.$$
Suppose that $a$, $b$ and $c$ satisfy the conditions in ($\gamma$) or
($\delta$) above. Then, it is easy to see that $a$, $b$ and $a+b-c$ do not
satisfy any of the conditions ($\alpha$)--($\delta$). Thus, the only
possibilities are ($\alpha$) and ($\beta$). We prove by induction on $i$ that
($\beta$) is impossible.

If $i=1$, we know by 7.4 that $k=j+1$ (since $i\le j\le k$) and that
$b-a=\frac12(j+1)$, so ($\beta$) is impossible in this case.  Assume that
($\alpha$) is the only possibility for for $i-1$. If $(\beta)$ holds for $i$,
we see from 7.4 that
$${\text{Hom}}_\y(V_{a+n-1-\frac12(j-i+1)}(\lambda_{j-1}),
V_{a+n-\frac32}(\lambda_{i-1})\ot V_c(\lambda_k))\ne 0,$$
or equivalently that
$${\text{Hom}}_\y(V_{a-\frac12}(\lambda_{i-1})\ot
V_{a+n-1-\frac12(j-i+1)}(\lambda_{j-1}), V_c(\lambda_k))\ne 0.$$
Since $(\alpha)$  holds for $i-1$, we get
 $$a+n-1-\frac12(j-i+1)-\left(a-\frac12\right)=\frac12(i+j-2), $$
i.e. $j=n$, contradicting our assumption that $j<n$. This completes the
induction, and proves that (30) implies ($\alpha$).

We now show, again by induction on $i$, that (30) implies that either (i) or
(ii) in the statement of 7.1 must hold. If $i=1$, the result follows from 7.4.

Assume the result for $i-1$. To complete the induction we consider four cases:
\vskip6pt\non
{\it Case 1.} $j<k\le n-2$. By 7.4, ($\alpha$) gives $k=i+j$ or $2n-i-j-2$,
$${\text{Hom}}_\y(V_{a-\frac12}(\lambda_{i-1})\ot
V_{a+\frac12(i+j-1)}(\lambda_{j+1}), V_c(\lambda_k))\ne 0,$$
and
$$c-\left(a-\frac12\right)=\frac12(j+1),$$
i.e. $c-a=\frac12 j$. Thus, either (i) or (ii) in 7.1 must hold for $i$.
\vskip6pt\non
{\it Case 2.} $j=k<n-2$. Again,  ($\alpha$) gives
$${\text{Hom}}_\y(V_{a-\frac12}(\lambda_{i-1})\ot
V_{a+\frac12(i+j-1)}(\lambda_{j+1}), V_c(\lambda_j))\ne 0.$$
Unfortunately, 7.1 does not apply to this because $j+1\nleq j$. However, the
non-vanishing of this last space of homomorphisms is equivalent to
$${\text{Hom}}_\y(V_{c-n+1}(\lambda_{j})\ot V_{a-\frac12}(\lambda_{i-1}),
V_{a+\frac12(i+j-1)-n+1}(\lambda_{j+1}))\ne 0,$$
and hence, twisting by $\v$ and then applying $\tau_{n-1}$, to
$${\text{Hom}}_\y(V_{-a+\frac12}(\lambda_{i-1})\ot V_{-c+n-1}(\lambda_{j}),
V_{-a-\frac12(i+j-1)+n-1}(\lambda_{j+1}))\ne 0.$$
We can apply the induction hypothesis to this inequality, and this gives that
$j+1=i-1+j$ or $2n-(i-1)-j-2$, and
$$-c+n-1-\left(-a+\frac12\right)=\frac12(i-1+j)\ \text{and}\
-a-\frac12(i+j-1)+n-1-\left(-a+\frac12\right)=\frac12j.$$
In both cases, we get $i+2j=2n-2$ and $c-a=\frac12j$, so 7.1 (ii) is satisfied
(we already know that $b-a=\frac12(i+j)$).
\vskip6pt\non
{\it Case 3.} $i<n-2$, $j=k=n-2$. We  show  that this case is possible only if
$i=2$ (it is obvious that $i$ must be even).
We first determine the value of $c$. Observe that (30) implies that
$${\text{Hom}}_\y(V_{c-n+1}(\lambda_{n-2})\ot V_a(\lambda_i),
V_{b-n+1}(\lambda_{n-2}))\ne 0,$$
and hence, twisting by $\v$ and applying $\tau_{n-1}$, we get
$${\text{Hom}}_\y( V_{-a}(\lambda_i)\ot V_{-c+n-1}(\lambda_{n-2}),
V_{-b+n-1}(\lambda_{n-2}))\ne 0.$$
Since $(\alpha)$ must hold for this, we get
$$c=a+\frac12(n-i).$$
By 7.4, we see that (30) implies
$${\text{Hom}}_\y(V_{a-n+\frac12(i+3)}(\lambda_1)\ot
V_{a+\frac12}(\lambda_{i+1})\ot V_{a+\frac12(n+i-2)}(\lambda_{n-2}),
V_{a+\frac12(n-i)}(\lambda_{n-2}))\ne 0,$$
which gives
$${\text{Hom}}_\y( V_{a+\frac12}(\lambda_{i+1})\ot
V_{a+\frac12(n+i-2)}(\lambda_{n-2}), V_{a+\frac12(i+1)}(\lambda_1)\ot
V_{a+\frac12(n-i)}(\lambda_{n-2}))\ne 0.$$
By 7.4 again, the module on the right-hand side of this space of homomorphisms
is irreducible. Thus, if $v_1$ and $v_{n-2}$ are $\y$-highest weight vectors in
  $V_{a+\frac12(i+1)}(\lambda_1)$ and $V_{a+\frac12(n-i)}(\lambda_{n-2})$,
respectively, $v_1\ot v_{n-2}$ must be in the image of any non--zero
homomorphism $F$ in
$${\text{Hom}}_\y( V_{a+\frac12}(\lambda_{i+1})\ot
V_{a+\frac12(n+i-2)}(\lambda_{n-2}), V_{a+\frac12(i+1)}(\lambda_1)\ot
V_{a+\frac12(n-i)}(\lambda_{n-2})).$$
Since $\lambda_{i+1}-\lambda_1\in Q_J^+$, where $J=\{2,3,\ldots ,n\}$, we see
from 5.5 that
$${\text{Hom}}_{\uqgj}(V_{a+\frac12}(\lambda_{i})\ot
V_{a+\frac12(n+i-2)}(\lambda_{n-3}), V_{a+\frac12(n-i)}(\lambda_{n-3}))\ne 0.$$
The induction hypothesis on $n$ now proves that $i=2$.
\vskip6pt\non
{\it Case 4.} $i=j=k=n-2$. In this case, we have $b=a+n-2$ and
$c=a+\frac12(n-2)$. Equation (30) implies that
$${\text{Hom}}_\y(V_{c-n+1}(\lambda_{n-2})\ot V_a(\lambda_{n-2}),
V_{b-n+1}(\lambda_{n-2}))\ne 0.$$
But ($\alpha$) does not hold for this if $n\ne 4$, and 7.1(ii) holds if $n=4$.

Finally, the inductive step, and with it the proof of the \lq only if\rq\ part
of 7.1, is complete.

We now prove the  \lq if\rq\ part. Suppose that (i) holds. Taking
$J=\{1,2,\ldots, n-2\}$, we see that $V_{a+\frac12(i+j)}(\lambda_j)\ot
V_{\frac12{a}}(\lambda_i)$ has a $\uqg$-highest weight vector of weight
$\lambda_{i+j}$. This vector cannot generate an irreducible highest weight
$\y$-submodule, since otherwise it would have an irreducible $\y$-submodule,
which would necessarily be of the form $V_c(\lambda_r)$ for some $r<i+j$, and
this is impossible by the \lq only if\rq\ part of 7.1.

Suppose now that (ii) holds. Recall from the discussion in Section 4 that we
may assume that $i\le j\le k$.

Consider first the case when  $n$, $i$ and $j$ are all even. By 7.5,
$$\align
{\roman{Hom}}_\y(V_{a-\frac12(n+i-1)}(\lambda_n) & \ot V_a(\lambda_i),
V_{a-\frac12(n-i-1)}(\lambda_n))\ne 0,\\
{\roman{Hom}}_\y(V_{a-\frac12(n-i-1)}(\lambda_n), &
V_{a+j-\frac12(n+i-1)}(\lambda_n)\ot V_{a+\frac12(i+j)-n+1}(\lambda_j))\ne 0.
\endalign$$
Hence,
$${\text{Hom}}_\y( V_{a-\frac12(n+i-1)}(\lambda_n)\ot V_a(\lambda_i),
V_{a+j-\frac12(n-i-1)}(\lambda_n)\ot  V_{a+\frac12(i+j)-n+1}(\lambda_j))\ne
0,$$
or equivalently,
$${\text{Hom}}_\y( V_a(\lambda_i)\ot V_{a+\frac12(i+j)}(\lambda_j),
V_{a+\frac12(n-i-1)}(\lambda_n)\ot V_{a+j-\frac12(n-i-1)}(\lambda_n))\ne 0.$$
We consider the composite of a non-zero element $F$ of this space homomorphisms
with the non-zero homomorphism
$$V_{a+\frac12(n-i-1)}(\lambda_n)\ot V_{a+j-\frac12(n-i-1)}(\lambda_n)\to
V_{a+\frac12j}(\lambda_{2n-2-i-j})$$
given by 7.5. This composite cannot be zero, otherwise the image of $F$ would
be a $\uqg$-module $N$ for which
$${\text{Hom}}_\y(^tN\ot V_a(\lambda_i), V_{a+\frac12(i+j)-n+1}(\lambda_j))\ne
0.\tag31$$
Moreover, the irreducible $\ung$-modules occurring in $N$ would be among the
set $$\{W(2\lambda_n), W(\lambda_{n-2}), W(\lambda_{n-4}),\ldots ,
W(\lambda_{2n-i-j}) \}.$$
Since $i\le k=2n-i-j-2$,
$$m_0(^tN\ot W(\lambda_i)) =0,\ \ \ \ m_0(W(\lambda_j)) =1,$$
so $(31)$ is impossible.

The proofs in the other cases are similar applications of 7.5 and the
$\ung$-module decomposition of the fundamental $\y$-modules. We omit the
details.

This completes the proof of the \lq if\rq\ part of 7.1.\qed\enddemo

It remains to give the
\demo{Proof of 7.2} We proceed by induction on $n$. As usual, the induction
starts at $n=3$, where the result is known from 6.2. Assume now that 7.2 is
known when $\ung$ is of type $D_m$ for $m<n$. To prove the result when $\ung$
is of type $D_n$, we consider first the case when $i\le j\le n-2$, and proceed
by induction on ${\text{min}}(i,n-j)$. The induction starts when $i=1$, this
case being covered by 7.4.

Assume that $V_a(\lambda_i)\ot V_b(\lambda_j)$ is not $\y$-highest weight. Let
$v_i$ and $v_j$ be $\y$-highest weight vectors in $V(\l_i)$ and $V(\l_j)$,
respectively, and let $N$ be an irreducible $\y$-quotient of $V(\l_i)\ot
V(\l_j)/\y.(v_i\ot v_j)$. Then, we have
$${\text{Hom}}_\y(V_a(\lambda_i)\ot V_b(\lambda_j),  N)\ne 0,$$
and $N_{\lambda_i+\lambda_j}=0$. Assume for a contradiction that  $b-a$ takes
none of the values
$$\align \frac12(j-i)+r,\ \ \ \ & 0<r\le{\text{min}}(i,n-j),\\
\noalign{or}\ \ n-1+\frac12(i-j)-r,\ \ \ \ & 0\le
r<{\text{min}}(i,n-j).\endalign$$
If $i+j\le n$, we use 7.4 to get
$${\text{Hom}}_\y(V_{a+\frac12(1-i)}(\lambda_1)\ot
V_{a+\frac12}(\lambda_{i-1})\ot V_b(\lambda_j), N)\ne 0,$$
and hence
$${\text{Hom}}_\y(V_{a+\frac12}(\lambda_{i-1})\ot V_b(\lambda_j),
V_{a+n-\frac12(i+1)}(\lambda_1)\ot N)\ne 0.\tag32$$
By the assumption on $b-a$ and the induction hypothesis on
${\roman{min}}(i,n-j)$, $V_a(\lambda_{i-1})\ot V_b(\lambda_j)$
is $\y$-highest weight. Hence, $\lambda_1+\lambda\ge \lambda_{i-1}+\lambda_j$,
so $\lambda=\lambda_i+\lambda_j-\eta$, where $\eta=\sum_{i=1}^nr_i\alpha_i\in
Q^+$ satisfies
$$\eta\le\alpha_1+2\alpha_2+\cdots+\alpha_{i-1}+2(\alpha_i+\cdots+\alpha_{n-2})+\alpha_{n-1}+\alpha_n.$$
If $r_1=0$, let $J=\{2,3,\ldots ,n\}$. Then, by 5.4,
$${\text{Hom}}_{\uqgj}(V_a(\lambda_{i-1})\ot V_b(\lambda_{j-1}), N_J)\ne 0.$$
By the induction hypothesis on $n$, we have
$$b-a=\cases\frac12(j-i)+r\ \ \text{for some}\ \ 0<r\le i-1,\ {\text{or}}\\
n-2+\frac12(i-j)-r\ \ \text{for some}\ \ 0\le r<i-1.\endcases$$
But all these values have beeen excluded.

Using $\lambda\in P^+$, one sees that the case $r_1>0$ is possible only if
$i=2$, and then either $\eta=\lambda_2$, or $\eta\in Q_{J'}^+$, where
$J'=\{1,2,\ldots ,n-2,n-1\}$ or $\{1,2,\ldots ,n-2,n\}$. In the first case,
$\lambda=\lambda_j$ and (32) becomes
$${\text{Hom}}_\y(V_{a+\frac12}(\lambda_1)\ot V_b(\lambda_j),
V_{a+n-\frac12(i+1)}(\lambda_1)\ot V_c(\lambda_j))\ne 0,\tag33$$
for some $c\in\Bbb C$. By the assumption on $b-a$,
$V_{a+\frac12}(\lambda_1)\ot V_b(\lambda_j)$ is $\y$-highest weight, but then
(33) contradicts 3.2. In the second case, 5.4 gives
$${\text{Hom}}_{Y(\ung_{J'})}(V_a(\lambda_2)\ot V_b(\lambda_j), N_{J'})\ne 0,$$
and then 6.2 gives
$$b-a=\frac12j\ \  {\text{or}}\ \ \frac12j+1.$$
Both of these values have been excluded.

Thus, we have obtained the desired contradiction when $i+j\le n$. If $i+j>n$,
one uses  a similar argument, but using 7.4 to replace (32) by
$${\text{Hom}}_\y(V_a(\lambda_i)\ot V_{b-\frac12}(\lambda_{j+1}), N\ot
V_{b+\frac12(1-j)}(\lambda_1))\ne 0.$$
This proves the \lq only if\rq\ part of 7.2 when $j\le n-2$.

For the \lq if\rq\ part, note that the $i=1$ case is contained in 7.4. Suppose
that $i>1$. If $b-a=\frac12(j-i)+r$, where $0<r\le{\text{min}}(i,n-j)$, let
$J''=\{1,2,\ldots ,n-1\}$. By 6.2, $V_a(\lambda_i)_{J''}\ot
V_b(\lambda_j)_{J''}$ is not $\y$-highest weight, so by 5.5, neither is
$V_a(\lambda_i)\ot V_b(\lambda_j)$. If $b-a=n-i+\frac12(i-j)-r$, where
$0<r<{\text{min}}(i,n-j)$, one uses the same argument with $J''$ replaced by
$\{2,3,\ldots ,n\}$ and uses the induction hypothesis on $n$ instead of 6.2...
For the remaining value $b-a=n-1+\frac12(i-j)$, note that, if $i\ne j$,  we
have
$${\text{Hom}}_{\y}(V_a(\lambda_i)\ot V_{a+n-1+\frac12(i-j)}(\lambda_{j}),
V_{a+n-1-\frac12(j)}(\lambda_{j-i}))\ne 0,$$
by 7.1, while if $i=j$, then by 3.4 and 3.6, we have
$${\roman{Hom}}_\y(V_a(\lambda_i)\ot V_{a+n-1}(\lambda_i),\Bbb C)\ne 0.$$
In any case, this implies that $V_a(\lambda_i)\ot V_b(\lambda_j)$ is not
$\y$-highest weight.

We now consider part (ii). If $j=n$, the result follows by the above argument,
using 5.5 and 6.2 (and the same $J''$). If $j=n-1$, replace $J''$ with
$\{1,2,\ldots ,n-2,n\}$.

Finally, parts (iii) and (iv) follow immediately from 7.5.
\qed\enddemo

\vskip24pt\centerline{\bf 8. The $B_n$ and $C_n$ cases}
\vskip12pt\non
In this section, we give the analogues of Theorems 6.1 and 7.1 when $\ung$ is
of type $B_n$ or $C_n$.

\proclaim{Theorem 8.1} Let $\ung$ be of type $B_n$, and let $1\le i\le j\le
k\le n$, $a$, $b$, $c\in\Bbb C$. Then,
$${\roman{Hom}}_\y(V_a(\l_i)\ot V_b(\l_j),V_c(\l_k))\ne 0$$
if and only if one of the following holds:
\vskip6pt\non
(i) $i+j\le n-1$, $k=i+j$, $b-a=i+j$, $c-a=j$;

\non(ii) $i<n$, $j=k=n$, $b-a=n+i-1$, $c-a=n-i-1$. \qed\endproclaim

The proof of this theorem is very similar to that of 7.1. Although we shall
omit the details, we remark that the argument used to prove the existence of a
non-zero homomorphism of $\y$-modules
$$V_a(\l_1)\ot V_b(\l_j)\to V_c(\l_{2n-2-i-j})$$
(for suitable $a,b,c$) when $\ung$ is of type $D_n$ fails to produce a non-zero
homomorphism
$$V_a(\l_i)\ot V_b(\l_j)\to V_c(\l_{2n-i-j})$$
when $\ung$ is of type $B_n$ (as would be predicted by condition (D) alone)
because of a difference in the way that tensor products of spin modules for
$\ung$ behave in the two cases. Namely, in the $B_n$ case, the tensor product
of the spin module with itself contains every fundamental $\ung$-module except
the spin module,  whereas in the $D_n$ case, the tensor product of the two spin
modules, or of a spin module with itself, contains only \lq half\rq\ the
remaining fundamental $\ung$-modules.

\proclaim{Theorem 8.2}  Let $\ung$ be of type $C_n$, and let $1\le i\le j\le
k\le n$, $a$, $b$, $c\in\Bbb C$. Then,
$${\roman{Hom}}_\y(V_a(\l_i)\ot V_b(\l_j)\ot V_c(\l_k),\Bbb C)\ne 0$$
if and only if $i+j\le n$, $k=i+j$, $b-a=\frac12(i+j)$,
$c-a=\frac12 j$.\qed\endproclaim

The proof of this theorem is similar to that of 6.1. Note that the fundamental
$\y$-modules are irreducible as $\ung$-modules in both the $A_n$ and $C_n$
cases (see [11] and [8]).
\vskip12pt\non{\it Remark.} We can use 6.1, 7.1, 8.1 and 8.2 to prove 4.2,
avoiding the use of the PRV conjecture. In fact, we can prove a more general
result. Suppose, for example, that $\ung$ is of type $D_n$ and that $i,j\le
n-2$, $a,b\in\Bbb C$. Let $V$ be any irreducible quotient $\y$-module of
$V_a(\l_i)\ot V_b(\l_j)$, and let $\l$ be the highest weight of $V$ as a
$\ung$-module. Then,
$${\roman{Hom}}_\ung(W(\l_i)\ot W(\l_j),W(\l))\ne 0.\tag34$$
Indeed, since $\l$ is dominant and $\le \l_i+\l_j$, we have either
\vskip6pt\non(i) $\l=\l_k$ for some $k$, or

\non(ii) $\l=\l_k+\l_\ell$ for some $k$, $\ell$.
\vskip6pt\non In case (i), we know that $(i,j,k)$ satisfies the conditions in
4.5, and then (34) is easily checked. In case (ii), $\l_i+\l_j-\l_k-\l_\ell\in
Q_J^+$, where $J=\{2,3,\ldots,n\}$, so by 5.4 and an obvious induction on $n$,
we have
$${\roman{Hom}}_{\ung_J}(W(\l_i)_J\ot W(\l_j)_J,W(\l)_J)\ne 0,$$
which implies
$${\roman{Hom}}_{\ung}(W(\l_i)\ot W(\l_j),W(\l))\ne 0.$$
Similar arguments apply in the other cases.

\bigpagebreak

\vskip36pt\centerline{\bf 9. The quantum affine case}
\vskip12pt\non In this section, we indicate how to translate the preceding
results from the context of Yangians to that of quantum affine algebras. We use
freely the notation established in [8], Chapter 12. We assume throughout that
the deformation parameter $\epsilon$ is not a root of unity.

To find the quantum affine version of 6.1, for example, one replaces $\y$ by
$U_\epsilon(\hat\ung)$, $V_a(\l_i)$ by $V_\epsilon(\l_i,a)$, etc., and
conditions such as
$$b-a=\frac12(i+j),\ \ \ c-a=\frac12 j$$
in 6.1 (i) by
$$b/a=\epsilon^{i+j},\ \ \ c/a=\epsilon^j.$$
Similarly, in 6.2, the condition for $V_\epsilon(\l_i,a)\ot V_\epsilon(\l_j,b)$
not to be a highest weight $U_\epsilon(\hat\ung)$-module is
$$b/a=\epsilon^{j-i+2r}\ \ \ \text{for some $0<r\le{\roman{min}}(i,n+1-j)$.}$$
The main results in Sections 7 and 8 can be translated in the same way. We
leave this to the reader, as well as the straightforward problem of
appropriately reformulating the proofs.

\vskip36pt\centerline{\bf 10. Appendix: Dorey's rule and affine Toda theories}
\vskip12pt\noindent In this section, we sketch how Dorey's condition arises in
the context of ATFTs. We shall consider only those ATFTs based on untwisted
affine algebras.

We begin by summarizing some results related to Coxeter elements, for which we
follow [15]. Let $\alpha_0=-\theta$, $X_0^\pm =X_\theta^\mp$, and $\hat
I=I\amalg\{0\}$. Let $k_i$ ($i\in I$) be the coprime positive integers such
that
$$\sum_{i\in\hat I}k_i\alpha_i=0$$
(so that $k_0=1$), and set
$$X^\pm =\sum_{i\in\hat I}\sqrt{k_i}X_i^\pm .$$
Since $X^+$ is a regular element [15], its centralizer is a Cartan subalgebra
$\unh'$ of $\ung$. Note that $[X^+,X^-]=0$, so $X^\pm \in\unh'$. Recall also
that $h=\sum_{i\in\hat{I}}k_i$.

Let $H\in\unh$  be such that $\alpha_i(H)=1$ for all $i\in I$, and set
$$A={\roman{exp}}\left(\frac{2\pi\sqrt{-1}}h H\right).$$
Thus, $A$ lies in a connected complex simple Lie group $G$ with Lie algebra
$\ung$.

Note that the centralizer of $A$ in $\ung$ is $\unh$. On the other hand, it is
clear that
$${\roman{Ad}}(A)(X^\pm)=\omega^{\pm 1}X^\pm,$$
where $\omega=e^{2\pi\sqrt{-1}/h}$. It follows that
${\roman{Ad}}(A)(\unh')=\unh'$. In fact, it is known [15] that
${\roman{Ad}}(A)\vert_{\unh'}$ is a Coxeter transformation of $\unh'$, i.e. one
can choose an ordered set of simple roots
$\alpha_1',\alpha_2',\ldots,\alpha_n'$ of $\ung$ with respect to $\unh'$ such
that
$$\g'\equiv {\roman{Ad}}(A)\vert_{\unh'}=s_1's_2'\ldots s_n',$$
where $s_i'$ is the $i$th simple reflection in the Weyl group of $\ung$ with
respect to $\unh'$. Define, for $i\in I$,
$$\phi_i'=s_n's_{n-1}'\ldots s_{i+1}'(\alpha_i'),$$
and let $R_i'$ be the $\g'$-orbit of $\phi_i'$.

Choose root vectors $X_{\alpha'}$, for every root $\alpha'$ of $\ung$ with
respect to $\unh'$, such that
$${\roman{Ad}}(A)(X_{\alpha'})=X_{\g'(\alpha')},$$
and set
$$\tilde{H}_i=\sum_{\beta'\in R_i'}X_{\beta'}.\tag35$$
It is clear that ${\roman{Ad}}(A)(\tilde{H}_i)=\tilde{H}_i$, so
$\tilde{H}_i\in\unh$. Obviously, the $\tilde{H}_i$ are linearly independent,
and hence form a basis of $\unh$.

\vskip12pt We now turn to Toda field theory. The ATFT based on the affine Lie
algebra $\hat\ung$ is defined by the lagrangian
$${\Cal L}(\Psi)=\iint
\left\{\left\|\frac{\partial\Psi}{\partial t}\right\|^2\right.
-\left\|\frac{\partial\Psi}{\partial x}\right\|^2
\left.+\frac{m^2}{\beta^2}\left(e^{\roman{ad}(\Psi)}(X^+),X^-\right)\right\}dxdt,$$
where $\Psi$ is a function of the coordinates $(x,t)$ on $1+1$ dimensional
spacetime with values in $\unh$, $(\ ,\ )$ is the invariant bilinear form on
$\ung$, $m^2$ is a (positive) mass scale, and $\beta$ is a coupling constant
(usually either real or purely imaginary). Since the $\tilde{H}_i$ are a basis
of $\unh$, we can write
$$\Psi=\sum_{i\in I}\psi_i\tilde{H}_i,\tag36$$
where the $\psi_i$ are scalar-valued functions. The component $\psi_i$ is
associated with the $i$th particle of the theory. The potential term
$$V(\Psi)=\left(e^{{\roman{ad}}(\Psi)}(X^+),X^-\right)$$
in ${\Cal L}(\Psi)$ can be expanded formally as a power series in the $\psi_i$,
$$V(\Psi)=\sum_{p=0}^\infty\sum_{i_1,i_2,\ldots, i_p\in I}V_{i_1i_2\ldots
i_p}\psi_{i_1}\psi_{i_2}\ldots\psi_{i_p},$$
and one says that there is a coupling (or fusing) between the particles
labelled $i_1,i_2,\ldots,i_p$ if $V_{i_1i_2\ldots i_p}\ne 0$.

 From (36),
$$V_{i_1i_2\ldots i_p}=
\frac1{p!}\sum_\pi\left([\tilde{H}_{i_{\pi(1)}},[\tilde{H}_{i_{\pi(2)}},\ldots,[\tilde{H}_{i_{\pi(p)}},X^+]\cdots]],X^-\right),$$
where the sum is over all permutations $\pi$ of $\{1,2,\ldots,p\}$. Next, using
(35), we get
$$V_{i_1i_2\ldots i_r}=\frac1{p!}\sum_\pi\sum_{\beta_1'\in R_{i_{\pi(1)}}'}
\sum_{\beta_2'\in R_{i_{\pi(2)}}'}\cdots\sum_{\beta_p'\in R_{i_{\pi(p)}}'}
\left([X_{\beta_1'},[X_{\beta_2'},\ldots,[X_{\beta_p'},X^+]\cdots]],X^-
\right).\tag37$$
Since the weight of
$$[X_{\beta_1'},[X_{\beta_2'},\ldots,[X_{\beta_p'},X^+]\cdots]]$$
with respect to $\unh'$ is $\beta_1'+\beta_2'+\cdots+\beta_p'$, it is clear
that the term on the right-hand side of (37) corresponding to
$\beta_1',\beta_2',\cdots,\beta_p'$ can be non-zero only if
$$\beta_1'+\beta_2'+\cdots+\beta_p'=0,$$
and hence that
$$V_{i_1i_2\ldots i_p}\ne 0\ \ \text{only if}\ \ 0\in
R_{i_1}'+R_{i_2}'+\cdots+R_{i_p}'.$$
Thus, the $p$-point coupling $V_{i_1i_2\ldots i_p}\ne 0$ only if
$i_1,i_2,\ldots,i_p$ satisfies (${\roman{D}}_p$). The converse statement also
holds when $p=3$, but this requires a case-by-case analysis, which we omit.

\vskip36pt\centerline{\bf 11. References}
\vskip12pt\non 1. D. Bernard, Hidden Yangians in 2D massive current algebras,
Commun. Math. Phys. {\bf 137} (1991), 191--208.

\non 2. D. Bernard and A. LeClair, Quantum group symmetries and non-local
currents in 2D QFT, Commun. Math. Phys. {\bf 142} (1991), 99--138.

\non 3. N. Bourbaki, El\'ements de Math\'ematique, Fasc. XXXIV, Groupes et
Algebres de Lie, Ch. IV--VI, Hermann, Paris, 1968.

\non 4. H. W. Braden, A note on affine Toda couplings, J. Phys. A {\bf 25}
(1992), L15--L20.

\non 5. H. W. Braden, E. Corrigan, P. E. Dorey and R. Sasaki, Affine Toda field
theory and exact S-matrices, Nucl. Phys.  {\bf B338} (1990), 689--746.

\non 6. V. Chari and A. N. Pressley, Yangians and R-matrices, L'Enseign. Math.
{\bf 36} (1990), 267--302.

\non 7. V. Chari and A. N. Pressley, Fundamental representations of Yangians
and singularities of R-matrices, J. reine angew. Math. {\bf 417} (1991),
87--128.

\non 8. V. Chari and A. N. Pressley, {\it A Guide to Quantum Groups}, Cambridge
University Press, Cambridge, 1994.

\non 9. G. W. Delius, M. T. Grisaru and D. Zanon, Exact S-matrices for
non-simply-laced affine Toda theories, Nucl. Phys. {\bf B382} (1992), 365--406.

\non 10. P. E. Dorey, Root systems and purely elastic S-matrices, Nucl. Phys.
{\bf B358} (1991), 654--676.

\non 11. V. Drinfel'd, Hopf algebras and the quantum Yang--Baxter equation,
Soviet Math. Dokl. {\bf 32} (1985), 254--258.

\non 12. V. Drinfel'd, A new realization of Yangians and quantum affine
algebras, Soviet Math. Dokl. {\bf 36} (1988), 212--216.

\non 13. L. D. Faddeev, Quantum completely integrable models in field theory,
Soviet Scientific Reviews Sect. C {\bf 1}, 107--155, Harwood Academic
Publishers, Chur, Switzerland.

\non 14. V. G. Kac, {\it Infinite dimensional Lie algebras}, Birkh\"auser,
Boston, 1983.

\non 15. B. Kostant, The principal three-dimensional subgroup and the Betti
numbers of a complex simple Lie group, Am. J. Math. {\bf 81} (1959), 973--1032.

\non 16. S. Kumar, A proof of the Parthasarathy--Ranga Rao--Varadarajan
conjecture, Invent. Math. {\bf 93} (1988), 117--130.

\non 17. N. J. MacKay, The full set of $c_n$-invariant factorized S-matrices,
J. Phys. A {\bf 25} (1992), L1343--L1349.

\non 18. O. Mathieu, Construction d'un groupe de Kac--Moody et applications,
Compositio Math. {\bf 69} (1989), 37--60.

\non 19. E. Ogievetsky, N. Yu. Reshetikhin and P. Wiegmann, The principal
chiral field in two dimensions on classical Lie algebras, Nucl. Phys. {\bf
B280} (1987), 45--96.

\non 20. D. I. Olive, N. Turok and J. W. R. Underwood, Affine Toda solitons and
vertex operators, Nucl. Phys. {\bf B409} (1993), 509--546.

\non 21. K. R. Parthasarathy, R. Ranga Rao and V. S. Varadarajan,
Representations of complex semi-simple Lie groups and Lie algebras, Ann. Math.
{\bf 85} (2) (1967), 383--429.

\non 22. T. A. Springer, Regular elements of finite reflection groups, Invent.
Math. {\bf 25} (1974), 159--198.

\non 23. R. Steinberg, Finite reflection groups, Trans. Am. Math. Soc. {\bf 91}
(1959), 493--504.
\vskip36pt
{\eightpoint{
$$\matrix\format\l&\l&\l&\l\\
\phantom{.} & {\text{Vyjayanthi Chari}}\phantom{xxxxxxxxxxxxx} & {\text{Andrew\
Pressley}}\\
\phantom{.}&{\text{Department of Mathematics}}\phantom{xxxxxxxxxxxxx} &
{\text{Department of Mathematics}}\\
\phantom{.}&{\text{University of California}}\phantom{xxxxxxxxxxxxx} &
{\text{King's College}}\\
\phantom{.}&{\text{Riverside}}\phantom{xxxxxxxxxxxxx} & {\text{Strand}}\\
\phantom{.}&{\text{CA 92521}}\phantom{xxxxxxxxxxxxx} & {\text{London WC2R
2LS}}\\
\phantom{.}&{\text{USA}}\phantom{xxxxxxxxxxxxx} & {\roman{UK}}\\
&{\text{email: chari\@ucrmath.ucr.edu}}\phantom{xxxxxxxxxxxxx} &
{\text{email:anp\@uk.ac.kcl.mth}}
\endmatrix$$
}}

\enddocument